\documentclass[a4paper,11pt]{article}
\usepackage{a4wide,amsmath,amssymb,bbm}
\usepackage[english]{babel}
\usepackage{hyperref,enumerate}
\usepackage{tikz} 
\hypersetup{
    colorlinks,
    linkcolor={black},
    citecolor={green!50!black},
    urlcolor={blue!50!black}
}

\newcommand{\Ad}{\mathrm{Ad}}
\newcommand{\ad}{\mathrm{ad}}
\newcommand{\Str}{\mathrm{Str}}
\newcommand{\alg}[1]{\mathfrak{#1}}

\def\be{\begin{equation}}
\def\ee{\end{equation}}
\newcommand{\STr}{\operatorname{Str}}
\newcommand{\op}{\widetilde{\mathcal{O}}}

\newcommand{\alert}[1]{{\color[rgb]{1,0,0} #1}}
\makeatletter

\@addtoreset{equation}{section} \makeatother

\begin{document}

\hfill{NORDITA 2017-065}

\vspace{30pt}

\begin{center}
{\LARGE{\bf On non-abelian T-duality and deformations of supercoset string sigma-models}}

\vspace{50pt}

Riccardo Borsato$\, ^a$ \ \ and \ \ Linus Wulff$\, ^b$

\vspace{15pt}

{
\small {$^a$\it Nordita, Stockholm University and KTH Royal Institute of Technology,\\ Roslagstullsbacken 23, SE-106 91 Stockholm, Sweden}\\
\small {$^b$\it Department of Theoretical Physics and Astrophysics, Masaryk University, 611 37 Brno, Czech Republic}
\\
\vspace{12pt}
\texttt{riccardo.borsato@su.se, linus.wulff@gmail.com}}\\

\vspace{100pt}

{\bf Abstract}
\end{center}
\noindent
We elaborate on the class of deformed T-dual (DTD) models obtained by first adding a topological term to the action of a supercoset sigma model and then performing (non-abelian) T-duality on a subalgebra $\tilde{\mathfrak{g}}$ of the superisometry algebra. 
These models inherit the classical integrability of the parent one, and they include as special cases the so-called homogeneous Yang-Baxter sigma models as well as their non-abelian T-duals. 
Many properties of DTD models have  simple algebraic interpretations. For example we show that their (non-abelian) T-duals---including certain deformations---are again in the same class, where $\tilde{\mathfrak{g}}$ gets enlarged or shrinks by adding or removing generators corresponding to the dualised isometries. 
Moreover, we show that Weyl invariance of these models is equivalent to $\tilde{\mathfrak{g}}$  being unimodular; when this property  is not satisfied one can always remove one generator to obtain a unimodular $\tilde{\mathfrak{g}}$, which is equivalent to (formal) T-duality. 
We also work out the target space superfields and, as a by-product, we prove the conjectured transformation law for Ramond-Ramond (RR) fields under bosonic non-abelian T-duality of supercosets, generalising it to cases involving also fermionic T-dualities.

\pagebreak 
\tableofcontents

\setcounter{page}{1}


\section{Introduction}
In this paper we investigate further the deformed T-dual (DTD) supercoset sigma models introduced in \cite{Borsato:2016pas}, and we find results that are of interest also when considering the undeformed case, i.e. when applying just non-abelian T-duality (NATD).
 
The construction of DTD models is equivalent to applying NATD on a centrally extended subalgebra as first suggested in \cite{Hoare:2016wsk}.\footnote{The first hint of the relation of YB models to NATD appeared in~\cite{Orlando:2016qqu} for the case of Jordanian deformations.}  The models are constructed by picking a subalgebra of the (super)isometry algebra $\tilde{\alg{g}}\subset \alg g$---the canonical example is the $AdS_5\times S^5$ superstring where $\alg g=\alg{psu}(2,2|4)$---and a 2-cocycle, i.e. an anti-symmetric linear map $\omega:\,\tilde{\mathfrak g}\otimes\tilde{\mathfrak g}\rightarrow\mathbbm{R}$ satisfying
\begin{equation}
\omega(X,[Y,Z])
+\omega(Z,[X,Y])
+\omega(Y,[Z,X])
=0\,,\qquad\forall X,Y,Z\in\tilde{\mathfrak g}\,.
\label{eq:2-cocycle}
\end{equation}
Together with an element of the corresponding group $\tilde g\in\tilde G$, the 2-cocycle defines a 2-form $B=\omega(\tilde g^{-1}d\tilde g,\tilde g^{-1}d\tilde g)$ which is closed, i.e. $dB=0$, thanks to the 2-cocycle condition. The idea behind the construction is to add this topological term to the supercoset sigma model Lagrangian and then perform NATD on $\tilde G$. If $\zeta B$ is added to the Lagrangian, with $\zeta$ a parameter, the resulting model can be thought of as a deformation of the non-abelian T-dual of the original model with deformation parameter $\zeta$. The classical integrability of the original sigma model is preserved by the deformation, since both adding a topological term and performing NATD preserve integrability. We refer to \cite{Borsato:2016pas} for more details on how this procedure relates to the construction of \cite{Hoare:2016wsk}.
Let us remark that DTD models may be constructed starting from a generic $\sigma$-model, for example the principal chiral model as in~\cite{Borsato:2016pas}, and the starting model does not have to be (classically) integrable. In this paper we will only consider the supercoset case.

It was proven in \cite{Borsato:2016pas} that the so-called Yang-Baxter (YB) sigma models \cite{Klimcik:2002zj,Klimcik:2008eq,Delduc:2013qra,Kawaguchi:2014qwa}, defined by an R-matrix solving the classical Yang-Baxter equation (CYBE), are equivalent to DTD models with invertible $\omega$. This relation was first conjectured and checked for many examples---in the language of T-duality on a centrally extended subalgebra---in \cite{Hoare:2016wsk}. See also \cite{Hoare:2016wca} for a more detailed discussion of some of the examples. In \cite{Borsato:2016pas} we used the fact that when $\omega$ is invertible its inverse $R=\omega^{-1}$ solves the CYBE, and therefore defines a corresponding YB model; by means of a field redefinition and relating the deformation parameters as $\eta=\zeta^{-1}$ we could prove the equivalence of the two sigma model actions~\cite{Borsato:2016pas}. 

Note that simply by setting the deformation parameter to zero, DTD models include all non-abelian and abelian T-duals of the original supercoset model, including fermionic T-dualities. Therefore all the statements we prove for DTD models apply also to (non-abelian) T-duals of supercoset models. They are also easily seen to describe all so-called TsT-transformations of the underlying supercoset model. In fact we will argue here that the class of DTD models is closed under the action of NATD, as well as certain deformations, meaning that applying these operations yields a new DTD model. They therefore represent a very broad class of integrable string sigma models.

\vspace{12pt}

It was shown in \cite{Borsato:2016pas} that these models are invariant under kappa symmetry, which is needed to interpret them as Green-Schwarz superstrings. From the results of \cite{Wulff:2016tju} it follows that their target spaces must solve the generalised supergravity equations of \cite{Arutyunov:2015mqj,Wulff:2016tju} that ensure the one-loop scale invariance of the string sigma model. To have a fully consistent superstring, however, we must require the stronger condition of Weyl invariance, which implies that the target space should be a solution of the more stringent standard supergravity equations. Here we show that Weyl invariance of the DTD model is equivalent to the Lie algebra $\tilde{\mathfrak g}$ being unimodular, i.e. its structure constants should satisfy $f_{ij}^j=0$. In fact, this condition is precisely the one found in \cite{Alvarez:1994np,Elitzur:1994ri} when analysing the Weyl invariance of bosonic sigma models under NATD by path integral considerations. The presence of $\omega$ and the deformation does not modify the supergravity condition.
When $\omega$ is invertible the condition is also equivalent to unimodularity of the R-matrix $R=\omega^{-1}$, as defined in \cite{Borsato:2016ose}, which was shown there to be the condition for Weyl invariance of YB models. The fact that these conditions are the same was in fact an important hint that the latter should have an interpretation involving NATD \cite{Hoare:2016wsk}.

Here we give the detailed proof of kappa symmetry for DTD models and extract the target space superfields from components of the torsion as was done for $\eta$ (i.e. YB) and $\lambda$ models in \cite{Borsato:2016ose}. In particular, the RR fields and dilaton are difficult to extract by other means but we find that they are given by the simple expressions
\begin{equation}
e^{-2\phi}=\mathrm{sdet}'\op\,,\qquad
\mathcal S^{\alpha1\beta2}=
-8i[\Ad_h(1+4\Ad_f^{-1}\op^{-T}\Ad_f)]^{\alpha1}{}_{\gamma1}\widehat{\mathcal K}^{\gamma1\beta2}\,,
\end{equation}
with $\op$ defined in (\ref{eq:def-op}) and $\mathcal S$ defined in (\ref{eq:calS})---for definitions of the remaining quantities see sections \ref{sec:DTD} and \ref{sec:target}. A by-product of these expressions is a formula for the transformation of RR fields under NATD for the case of supercosets. As we show in section \ref{sec:target} it agrees, for bosonic T-dualities, with the formula conjectured in \cite{Sfetsos:2010uq}, see also~\cite{Lozano:2011kb}, but our formula is valid also when doing fermionic T-dualities.

An advantage of the formulation of DTD models is that many statements about the sigma model boil down to simple algebraic statements about the Lie algebra $\tilde{\mathfrak g}$. One example is the Weyl invariance condition already mentioned, while another concerns their transformation under NATD---possibly including additional deformation. The advantages are clear also when discussing the isometries of these models. We show that they fall into two classes; in fact, besides the standard ones, i.e. the unbroken part of the $G$ isometries, there are also certain (abelian) shift isometries. We prove that T-dualising on either type of isometry we get back a DTD model; in particular, T-dualising on the first type of isometries is equivalent to the simple operation of enlarging $\tilde{\mathfrak g}$ by the corresponding generators, while T-dualising on the shift isometries removes generators from $\tilde{\mathfrak g}$. The latter operation can be used to prove, in this context, that solutions of the generalised supergravity equations are (formally) T-dual to solutions of the standard supergravity equations \cite{Arutyunov:2015mqj}. For more general NATD, where one applies T-duality on both types of isometries at the same time, we propose that the resulting model is still obtained in a similar way, namely simply by adding to $\tilde{\mathfrak g}$ the isometry generators that lie outside of it and removing from it the generators that are inside. We show that this conjecture is indeed consistent, i.e. the resulting model is a well-defined DTD model, which turns out to be quite non-trivial. As already mentioned this suggests that the class of DTD models is closed under (bosonic and fermionic) NATD, including also the deformations considered here.

It was suggested in \cite{Borsato:2016pas} that it might be possible to think of all DTD models as non-abelian T-duals of YB models. Here we show that this is in fact not true by providing an example of a DTD model which cannot be obtained from a YB model by NATD.

The outline of the paper is as follows. In section \ref{sec:DTD} we introduce the DTD models based on supercosets, discuss their gauge invariances and the equivalence to YB models when $\omega$ is invertible. Section \ref{sec:symm} describes the two classes of global symmetries, or isometries, of these models. We also address the question of what happens if one performs NATD and deformation of a DTD model and argue that this gives a new DTD model, proving this in simpler cases. Models which cannot be obtained by NATD of YB models are also discussed. In section \ref{sec:kappa} we demonstrate the kappa symmetry of DTD models and write the DTD model as a Green-Schwarz superstring. Given these results it is then straightforward to derive the target space fields of the DTD model from components of the superspace torsion, which we do in section \ref{sec:target}. This includes a derivation of the Weyl-invariance condition for these models. In section \ref{sec:ex} we work out the supergravity background for two examples of DTD models. The first is equivalent to a well known TsT-background but is useful to demonstrate the procedure. The second example is one of the new examples which cannot be obtained from a YB model by NATD. We finish with some conclusions and open problems. Three appendices contain some useful algebraic identities, a derivation of the DTD model action and a proof of integrability.

\section{The Deformed T-dual models}\label{sec:DTD}
As described in the introduction the deformed T-dual (DTD) models are constructed as follows. We start with a supercoset sigma model, e.g. the $AdS_5\times S^5$ superstring \cite{Metsaev:1998it} or one of the other examples in \cite{Zarembo:2010sg,Wulff:2014kja}. We single out a subalgebra $\tilde{\mathfrak g}\subset\mathfrak g$ of the ($\mathbbm Z_4$-graded) superisometry algebra and write the group element as $g=\tilde gf$ with $\tilde g\in\tilde G$ and $f\in G$. This parametrization is of course redundant and introduces a corresponding $\tilde G$ gauge symmetry $\tilde g\rightarrow\tilde g\tilde h^{-1}$ and $f\rightarrow \tilde hf$ on which we will comment below. The second ingredient, which is responsible for the deformation, is a Lie algebra 2-cocycle $\omega$ on $\tilde{\mathfrak g}$ satisfying (\ref{eq:2-cocycle}). We add to the original supercoset sigma model action the term
\begin{equation}
S_\omega=\tfrac{T}{4}\int_\Sigma\,\zeta\omega(\tilde g^{-1}d\tilde g,\tilde g^{-1}d\tilde g)\,,
\end{equation}
where $\zeta$ is a parameter introduced to keep track of the deformation---if there exist many 2-cocycles we could introduce a parameter for each.\footnote{If $\omega$ has mixed Grassmann even-odd components the corresponding deformation parameter $\zeta$ would be fermionic. Since the interpretation of such a fermionic deformation is not so clear we will generally assume that $\omega$ has only even-even and odd-odd components and that $\zeta$ is real.} As explained already, this is equivalent to adding a B-field to the action, which is closed by virtue of the 2-cocycle condition. This term is therefore topological and has no effect on local properties of the theory---issues with boundary conditions are more subtle and will not be considered here. 
The final step is to perform NATD on $\tilde{\mathfrak g}$. This is done in the usual way by gauging the global $\tilde{\mathfrak g}$ symmetry and integrating out the gauge field. This procedure guarantees that properties like integrability are preserved, see appendix \ref{sec:integr} for an explicit proof. However, since T-duality is a non-local transformation of the fields of the sigma model, $\omega$ will now affect \emph{local} properties of the deformed model.

If $\omega$ is a coboundary, meaning that $\omega(X,Y)=f([X,Y])$ for some function $f:\,\tilde{\mathfrak g}\rightarrow\mathbbm{R}$, the B-field is exact; this is equivalent to no deformation at all since $B$ is pure gauge---alternatively a field redefinition can remove the $\zeta$ dependent contributions in the deformed model. Therefore non-trivial deformations are classified by the second (Lie algebra) cohomology group $H^2(\tilde{\mathfrak g})$. The same group also classifies non-trivial central extensions of $\tilde{\mathfrak g}$, consistent with the interpretation of these models as arising from NATD on a centrally extended subalgebra of the isometry algebra \cite{Hoare:2016wsk}.

Performing the above procedure one obtains the DTD supercoset model action
\begin{equation}
S=-\tfrac{T}{2}\int d^2\sigma\,\tfrac{\gamma^{ij}-\epsilon^{ij}}{2}\Str\big(J_i\hat d_fJ_j+(\partial_i\nu-\hat d_f^TJ_i)\op^{-1}(\partial_j\nu+\hat d_fJ_j)\big)\,,\qquad\gamma^{ij}=\sqrt{-h}h^{ij}\,,
\label{eq:S-DTD}
\end{equation}
and we refer to appendix~\ref{app:action} for the details of its derivation.
Here $J=dff^{-1}$ encodes the degrees of freedom in $f$, while $\nu\in\tilde{\mathfrak g}^*$ denotes the dualised degrees of freedom coming from $\tilde g$. We have further defined
\begin{equation}
\hat d_f=\Ad_f\hat d\Ad_f^{-1}\,,\qquad\hat d=P^{(1)}+2P^{(2)}-P^{(3)}\,,\quad\hat d^T=-P^{(1)}+2P^{(2)}+P^{(3)}\,,
\label{eq:dhat}
\end{equation}
where $P^{(i)}$ project onto the corresponding $\mathbbm{Z}_4$-graded component of $\mathfrak g=\sum_{i=0}^3\mathfrak g^{(i)}$ and $\op^{-1}$ is the inverse\footnote{Notice that $\op\op^{-1}=\tilde P^T$ and $\op^{-1}\op=\tilde P$ rather than $1$.} of the linear operator $\op:\tilde{\alg{g}}\to \tilde{\alg{g}}^*$
\begin{equation}\label{eq:def-op}
\op=\tilde P^T(\hat d_f-\ad_\nu-\zeta\omega)\tilde P\,.
\end{equation}
Given a basis $\{T_i\}$ of $\tilde{\mathfrak g}$ and using the fact that $\mathfrak g$ has a non-degenerate metric given by the supertrace, we define the Lie algebra $\tilde{\mathfrak g}^*\subset\mathfrak g$ dual to $\tilde{\mathfrak g}$ by taking as dual basis $\{T^i\}$, where $\Str(T^jT_i)=\delta^j_i$. Then we have $\tilde P$ and $\tilde P^T$ which are projectors onto $\tilde{\mathfrak g}$ and $\tilde{\mathfrak g}^*$ respectively. At the same time we are thinking of the 2-cocycle $\omega$ as a map $\omega:\,\tilde{\mathfrak g}\rightarrow\tilde{\mathfrak g}^*$ so that the cocycle condition takes the form
\begin{equation}
\omega[x,y]=\tilde P^T\left([\omega x,y]+[x,\omega y]\right)\,,\qquad\forall x,y\in\tilde{\mathfrak g}\,.
\label{eq:cocycle2}
\end{equation}
Therefore, modulo the projector on the right-hand-side, $\omega$ acts as a derivation with respect to the Lie bracket, similarly to $\ad_\nu$ which is a derivation thanks to the Jacobi identity.

In general one needs to make sure that the inverse $\op^{-1}$ exists in order to be able to define the model, and this puts some restrictions on the subalgebra $\tilde{\mathfrak g}$. By expanding in the parameter $\zeta$ we can think of the DTD model as a deformation of the non-abelian T-dual of the original model, since taking $\zeta=0$ reduces to ordinary NATD. Therefore, at least for a small deformation parameter the invertibility is guaranteed if one can apply NATD with respect to $\tilde{\mathfrak g}$.
There may also be cases in which NATD cannot be implemented but the operator is invertible for finite values of $\zeta$, i.e. the cocycle removes the 0-eigenvalues of $\op$.

We now want to turn to the discussion of the gauge invariances of the action~\eqref{eq:S-DTD} of DTD models. Besides the fermionic kappa symmetry, which will be discussed separately in section~\ref{sec:kappa}, the action has two types of gauge invariances:
\begin{itemize}
	\item[1.] \emph{Local Lorentz invariance}:
	\begin{equation}
	f\rightarrow fh\,,\qquad h\in H=G^{(0)}\,.
	\label{eq:LocalLorentz}
	\end{equation}
	\item[2.] \emph{Local $\tilde{G}$ invariance}:
	\begin{equation}
	f\rightarrow\tilde hf\,,\quad\nu\rightarrow\tilde P^T\left(\Ad_{\tilde h}\nu+\zeta\frac{1-e^{\ad_x}}{\ad_x}\omega x\right)\,,\qquad	\tilde h=e^x\in\tilde G\subset G\,.
	\label{eq:LocalGtilde}
	\end{equation}
\end{itemize}
The former is obvious and, as in the case of supercosets, it boils down to the fact that $P^{(0)}$ is missing in $\hat d$.
As mentioned at the beginning of this section, the latter comes about from the decomposition of the original group element as $g=\tilde gf$ where multiplication of $\tilde g$ from the right by an element of $\tilde G$ can be compensated for by multiplying $f$ on the left by the inverse group element.  To verify that the action is indeed invariant under the second type of symmetry we use the identities~\eqref{eq:transfo-op} and~\eqref{eq:transfo-dnu} that say how the transformations of $\op$ and $d\nu$ can be rewritten. Then the difference of the actions after and before the transformation~\eqref{eq:LocalGtilde} is proportional to
\begin{equation}
\int d^2\sigma\epsilon^{ij}\Str\Big(
2\partial_i\nu\tilde h^{-1}\partial_j\tilde h
+\tilde h^{-1}\partial_i\tilde h(\ad_\nu+\zeta\omega)(\tilde h^{-1}\partial_j\tilde h)
\Big)\,.
\end{equation}
The terms involving $\nu$ combine to a total derivative, and the one with $\omega$ is closed as already remarked, meaning that it is also a total derivative at least locally. This establishes the invariance of the action under the local transformation (\ref{eq:LocalGtilde}). This gauge invariance is obviously present also in the case of NATD, where the shift of $\nu$ is absent since $\zeta=0$.

\vspace{12pt}

The classical integrability of DTD models may be argued by the fact that they are obtained by adding a closed $B$-field and then applying NATD to the action of a supercoset, since neither of these operations breaks classical integrability, see e.g. \cite{Sfetsos:2013wia} for the argument in the case of NATD.
In appendix~\ref{sec:integr} we give a direct proof of the classical integrability of these models by showing that, similarly to what was shown in the case of DTD of PCM in~\cite{Borsato:2016pas}, the on-shell equations can be recast into the flatness condition
\be
\epsilon^{ij}(\partial_i\mathcal L_j+\mathcal L_i\mathcal L_j)=0\,,
\ee 
for the Lax connection
\be
\mathcal L_i=A^{(0)}_i+z A^{(1)}_i+\frac{1}{2}\left(z^2+z^{-2}\right)A^{(2)}_i+\frac{1}{2}\gamma_{ij}\epsilon^{jk}\left(z^{-2}-z^2\right)A^{(2)}_i+z^{-1}A^{(3)}_i\,,
\ee
where $z$ is the spectral parameter, $A^i=A^i_++A^i_-$ and $A_\pm^i\equiv \Ad_f^{-1}(\tilde A_\pm^i+J_\pm^i)$, with $\tilde A^i_\pm$  given in~\eqref{eq:solApm}. See appendix~\ref{app:action} for our notation.
Notice that the presence of the Lax connection still implies that we have conserved charges corresponding to the full original $\alg g$ symmetry. However, in contrast to the case of supercosets, for DTD models one cannot argue any more that they are all local, see appendix~\ref{sec:integr}.

\subsection{Relation to Yang-Baxter sigma models}\label{sec:YB}
Given a DTD model with a cocycle $\omega$ which is non-degenerate on $\tilde{\alg{g}}$, we can show that the action can be recast into the one of a YB model via a field redefinition. This result was first presented in~\cite{Borsato:2016pas} and we collect here more details of the proof.

Given a non-degenerate $\omega$ we denote its inverse by $R=\omega^{-1}$.
From the cocycle condition for $\omega$ it follows that $R$ solves the CYBE on $\tilde{\alg{g}}^*$. 
Conversely any solution of the CYBE on $\alg g$ defines an invertible 2-cocycle on a subalgebra\footnote{This follows from the fact that the subspace on which $R$ is invertible must be a subalgebra due to the CYBE \cite{MR674005}. Since $\omega=R^{-1}$ is a 2-cocycle on this subalgebra the subalgebra is quasi-Frobenius. Note that these results are true also for non-semisimple algebras and superalgebras.} $\tilde{\alg{g}}$, which demonstrates the one-to-one correspondence between DTD models with invertible $\omega$ and YB sigma models based on an R-matrix solving the CYBE.  
The field redefinition that relates the two models is
\begin{equation}\label{eq:field-red-DTD-YB}
\nu=\zeta\tilde P^T\frac{1-\Ad_{\bar g}}{\ad_{Rx}}\omega Rx\,,\qquad\bar g=e^{Rx}\in\tilde G\,,
\end{equation}
with $x\in\tilde{\mathfrak g}^*$ so that $Rx\in\tilde{\mathfrak g}$. In fact, using the identities in (\ref{eq:dmu}) and (\ref{eq:admu}) we find
\begin{equation}
d\nu=\tilde P^T(\ad_\nu+\zeta\omega)(\bar g^{-1}d\bar g)\,,\qquad
\tilde P^T\ad_\nu\tilde P=\zeta\tilde P^T\Ad_{\bar g}^{-1}\omega \Ad_{\bar g}\tilde P-\zeta\omega\,,
\end{equation}
and the action (\ref{eq:S-DTD}) becomes, after a bit of algebra,
\begin{equation}
S=-\tfrac{T}{2}\int d^2\sigma\,\tfrac{\gamma^{ij}-\epsilon^{ij}}{2}\Str\Bigg(
g^{-1}\partial_ig\hat d\left(1-\frac{R_g\hat d}{R_g\hat d-\zeta}\right) g^{-1}\partial_jg
+\bar g^{-1}\partial_i\bar g(\ad_\nu+\zeta\omega)\bar g^{-1}\partial_j\bar g
\Bigg)\,,
\end{equation}
where we have defined $g=\bar gf$ and $R_g=\Ad_g^{-1}R\Ad_g$. The last term vanishes up to a total derivative and we are left precisely with the action of the YB sigma model~\cite{Delduc:2013qra,Kawaguchi:2014qwa}
\begin{equation}
S=-\tfrac{T}{2}\int d^2\sigma\,\tfrac{\gamma^{ij}-\epsilon^{ij}}{2}\Str\left(g^{-1}\partial_ig\ \hat d\ (1-\eta R_g\hat d)^{-1}(g^{-1}\partial_jg)\right)\,,
\end{equation}
with deformation parameter $\eta=\zeta^{-1}$. 
In the special case when $\tilde{\mathfrak g}$ is abelian the DTD model is equivalent to a TsT transformation of the original supercoset sigma model, in agreement with the YB side for abelian $R$~\cite{Osten:2016dvf,Hoare:2016wsk}.

Let us mention that one can also construct a YB model for an R-matrix solving the modified CYBE, whose action takes essentially the same form as the above one \cite{Delduc:2013qra}; however, in that case it is not clear how to define the operator corresponding to $\omega$, and the relation to DTD models remains unclear. This case should be related by Poisson-Lie T-duality to the $\lambda$-model of~\cite{Sfetsos:2013wia,Hollowood:2014qma}.

We will argue in the next section that all (bosonic and fermionic) non-abelian T-duals of YB sigma models can be described as DTD models with certain degenerate $\omega$.
The converse is not true, in fact it is possible to identify DTD models which are not related to YB models by NATD; we refer to section~\ref{sec:not-YB} for an example and a discussion on this.

\section{Global symmetries}\label{sec:symm}
We will now describe the global symmetries, i.e. superisometries, of DTD models. Setting $\zeta=0$ and ignoring the presence of $\omega$ this discussion reduces to what one would have in the case of NATD.
In order to identify the global symmetries of these models we study the global transformations that leave the action invariant, \emph{modulo} gauge transformations with a global parameter, since the latter would not produce any Noether charge.
We find two types of global symmetries:\footnote{The two sets of transformations do not commute and their commutator is a transformation of the second type.}
\begin{itemize}
	\item[1.] \emph{Unbroken global $G$-transformations}:
	\begin{equation}\label{eq:GlobalG}
	\begin{aligned}
	&f\rightarrow g_0f\,,\quad\nu\rightarrow\tilde P^T\Ad_{g_0}\nu\,,\qquad g_0\in G \mbox{ and } g_0\notin\tilde G,\\
	&\mbox{such that}\quad (1-\tilde P)\Ad_{g_0}\tilde P=0\,,\quad {\tilde P^T\Ad_{g_0}^{-1}\omega\Ad_{g_0}\tilde P=\omega}\,.
	\end{aligned}
	\end{equation}
	The requirement $g_0\notin\tilde G$ comes from the fact that for $g_0\in\tilde G$ a combination of this isometry and the shift isometries described below is equivalent to a global $\tilde G$ gauge transformation.
	\item[2.] \emph{Global shifts of $\nu$}:
	\begin{equation}
	\nu\rightarrow\nu+\lambda\,,\qquad\lambda\in\tilde{\mathfrak g}^*\quad\mbox{such that}\quad\tilde P^T\ad_\lambda\tilde P=0
	\,.
	\label{eq:GlobalShift}
	\end{equation}
Note that the set of such $\lambda$'s will in general \emph{not} close into a subalgebra, although the corresponding isometry transformations of course commute since they are just shifts of $\nu$.
\end{itemize}
In the case when $\omega$ is invertible, which is equivalent to a YB sigma model with $R=\omega^{-1}$, it is not hard to show that these isometries coincide with the ones of the YB model which are normally written as $t\in\mathfrak g$ such that $R\ad_t=\ad_tR$.

Having global symmetries at our disposal means that we can gauge them and implement further NATD.  Before discussing the details of this in the next subsection,  we would like to exploit this possibility to make a comment regarding Weyl invariance of DTD models. As we prove in section~\ref{sec:target}, the target spaces of DTD models solve the standard supergravity equations if and only if the Lie algebra $\tilde{\alg{g}}$ is unimodular, i.e. $f_{ab}{}^b=0$. The standard supergravity equations are equivalent to the Weyl invariance at one-loop for the sigma-model, as opposed to just the scale invariance implied by the generalised supergravity equations \cite{Arutyunov:2015mqj,Wulff:2016tju}. In the non-unimodular case $f_{ab}{}^b\neq0$, and this defines a distinguished element of $\tilde{\mathfrak g}$; we can rotate the basis so that this element is $T_1$, i.e. $f_{1b}{}^b\neq0$ and $f_{ab}{}^b=0$ for $a\neq1$. The important observation is that the dual of the generator $T_1$ corresponds to an isometry. In fact, taking the trace of the Jacobi identity we find $f_{ab}{}^1=0$ and therefore
\begin{equation}
\Str(T_b\ad_{T^1}T_a)=f_{ab}{}^1=0\,,
\end{equation}
where $T^a \in\tilde{\mathfrak g}^*$. This confirms that $T^1$ satisfies (\ref{eq:GlobalShift}) and can be used to generate a shift isometry. Using the results of the next subsection, applying T-duality along the isometry direction $T^1$ one obtains a DTD model where $T_1$ is removed from $\tilde{\mathfrak g}$, so that the subalgebra that is left is now unimodular. Therefore, to each DTD model which is not Weyl invariant we can associate a Weyl invariant one obtained by (formal\footnote{Our discussion of isometries is at the level of the classical sigma model action, where the dilaton only appears in the combination $\mathcal F=e^\phi F$---together with RR fields---and in derivatives $\partial \phi$. When performing the T-duality we ignore the Fradkin-Tseytlin term, which will break the isometry referred to here.}) T-duality along a particular isometry direction. Obviously this possibility fails if there are obstructions to carrying out the T-duality, e.g. if the isometry in question is a null isometry. More generally, solutions of the \emph{generalised} supergravity equations are formally T-dual to solutions of the \emph{standard} supergravity equations ~\cite{Arutyunov:2015mqj,Wulff:2016tju}, and the above argument shows this relation in the specific context of DTD models.

\subsection{DTD of DTD models}
It is interesting to start from a DTD model as in~\eqref{eq:S-DTD} and further perform NATD, possibly including a deformation by a cocycle. We do this on the one hand to show that the application of these transformations on the sigma model does not require to start from a supercoset formulation, on the other hand to show that after these transformations we obtain a new DTD model. We will also use these results to argue that the example of the next subsection is not related to a YB model by NATD.

We can apply NATD by gauging the global isometries discussed above and dualising the corresponding directions.
Obviously, the choice of the type of isometries that we want to dualise will produce qualitative differences. 
In fact, if we consider isometries of the first type~\eqref{eq:GlobalG} and dualise a subalgebra $\hat{\mathfrak g}$, we essentially enlarge the subalgebra $\tilde{\mathfrak g}$. 
If instead we consider isometries of the shift type~\eqref{eq:GlobalShift} and dualise a subspace $\bar V^*\subset\tilde{\mathfrak g}^*$, then we  remove generators from the subalgebra $\tilde{\mathfrak g}$. The combination of isometry transformations that we consider here is therefore
\begin{equation}
f=\hat gf'\,,\quad\nu=\tilde P^T(\Ad_{\hat g}\nu'+\bar\lambda)\,,\qquad\mbox{with}\qquad\hat g\in\hat G\,,\quad\bar\lambda\in \bar V^*\,.
\end{equation}
After gauging them in the usual way we obtain a sigma model action which is just the one in~\eqref{eq:S-DTD}, where we replace\footnote{We will now use the notation $\check \nu\in\check{\alg{g}}$ for the field and the subalgebra of the DTD model from which we start. Similarly, we will denote the corresponding operators as $\check P,\ \check{\mathcal{O}}$, etc. We do this because we want to reserve the usual notation for the DTD model that is obtained at the end, after applying the further deformation of NATD.}
\begin{equation}
f\rightarrow f'\,,\qquad
J\rightarrow J'+\hat A\,,\qquad
d\nu\rightarrow d\check\nu+\check P^T[\hat A,\check\nu]+\bar a\,,
\end{equation}
where $\hat A\in \hat{\mathfrak g}$ is the non-abelian gauge field corresponding to the $\hat G$ isometries and $\bar a\in \bar V^*$ is the abelian gauge field corresponding to the shift isometries.
We add to the action the terms\footnote{For the sake of the discussion here we fix conformal gauge $\gamma^{+-}=\gamma^{-+}=\epsilon^{-+}=-\epsilon^{+-}=2$ where $\sigma^\pm= \tau\pm \sigma$. In principle it is also possible to add a deformation for the second type of isometry by adding a term $\bar a\bar\omega'\bar a$, but we will not consider this possibility further here.}
\begin{equation}
-T\int d^2\sigma\ \Str(\hat\nu \hat{F}_{+-}+\bar\rho \bar f_{+-}-\hat\zeta\hat A_+\hat\omega\hat A_-)\,,
\end{equation}
where $\hat{F}_{+-}=\partial_+\hat A_--\partial_-\hat A_++[\hat A_+,\hat A_-]$ and $\bar f_{+-}=\partial_+\bar a_--\partial_-\bar a_+$,  $\hat\nu$ and $\bar \rho$ are two new Lagrange multipliers, and $\hat\omega$ is a cocycle on $\hat{\alg{g}}$.
Integrating out $\hat\nu$ and $\bar\rho$ one obtains the action from which we started; to apply NATD we integrate out $\hat A$ and $\bar a$ instead. 

{We will now describe what happens when we dualise either $\hat{\alg{g}}$ or $\bar V^*$, and then use it to argue what should happen in the most general case where one dualises on both at the same time.}\footnote{In the rest of this section we absorb the parameter $\zeta$ into $\omega$ to simplify the expressions.}

\paragraph{Dualising type 1 isometries}
Consider first isometries of type 1 above, where we have $\hat P+\check P=\tilde P$ and $\hat P\check P=0$. After a bit of algebra and dropping primes, we find that the new action takes the form $S=-T\int d^2\sigma \mathrm{Str}(J_+\hat d_fJ_-+(\partial_+\nu-\hat d_f^TJ_+)\mathcal{Q}(\partial_-\nu+\hat d_fJ_-))$ where $\nu=\check\nu+\hat\nu$ and $\mathcal Q$ is an operator acting on $\tilde{\mathfrak g}=\check{\mathfrak g}\oplus\hat{\mathfrak g}$ which can be written in a $2\times 2$ block form as
\be
\mathcal{Q}=
\left(
\begin{array}{cc}
\check{\mathcal O}^{-1}+\check{\mathcal O}^{-1}(\hat d_f-\ad_{\check\nu})U^{-1}(\hat d_f-\ad_{\check\nu})\check{\mathcal O}^{-1} & -\check{\mathcal O}^{-1}(\hat d_f-\ad_{\check\nu})U^{-1}\\
-U^{-1}(\hat d_f-\ad_{\check\nu})\check{\mathcal O}^{-1} & U^{-1}
\end{array}
\right),
\ee
where\footnote{The operators $\check{\mathcal{O}},\hat{\mathcal{O}}$ are obtained from $\op$ by dressing $\nu, \omega$ and the projectors with checks or hats.} $U=\hat{\mathcal O}-\hat P^T(\hat d_f-\ad_{\check\nu})\check{\mathcal O}^{-1}(\hat d_f-\ad_{\check\nu})\hat P$.
It is straightforward to check that if we take $\omega=\check\omega+\hat\omega$ and define $\op$ as in~\eqref{eq:def-op}, then its decomposition in block form is
\be
\op
=
\left(
\begin{array}{cc}
\check{\mathcal O} & \check P^T(\hat d_f-\ad_{\check\nu})\hat P\\
\hat P^T(\hat d_f-\ad_{\check\nu})\check P & \hat{\mathcal O}
	\end{array}
\right),
\ee
and that $\mathcal Q=\op^{-1}$.
Therefore performing DTD by exploiting the unbroken isometries of the first type is equivalent to the simple operation of enlarging the dualised subalgebra as $\tilde{\mathfrak g}=\check{\mathfrak g}\oplus\hat{\mathfrak g}$, which is a Lie algebra due to the isometry condition $[\hat{\mathfrak g},\check{\mathfrak g}]\subset\check{\mathfrak g}$. As for the deformation, we are just adding new contributions, and $\omega=\check\omega+\hat\omega$ is a 2-cocycle on $\tilde{\mathfrak g}$ due to the isometry conditions in~\eqref{eq:GlobalG}.

\paragraph{Dualising type 2 isometries}

For isometries of type 2 we have $\bar P^T$ that projects on the space $\bar V^*$, so that  $\bar P\check P=\check P\bar P=\bar P$ and $\tilde P=\check P-\bar P$.
When integrating out $\bar a_\pm$ we get equations where $\bar P \check{\mathcal{O}}^{-1}$ appears, so that it is convenient to use the block decomposition on the space $\tilde{\alg{g}}\oplus \bar V$
\begin{align}
\check{\mathcal O}^{-1}
\equiv&
\left(
\begin{array}{cc}
\op & \tilde P^T(\hat d_f-\ad_{\tilde\nu}-\check\omega)\bar P\\
\bar P^T(\hat d_f-\ad_{\tilde\nu}-\check\omega)\tilde P & \bar P^T(\hat d_f-\ad_{\tilde\nu}-\check\omega)\bar P
	\end{array}
\right)^{-1}
\\
=&
\left(
\begin{array}{cc}
\op^{-1}+\op^{-1}(\hat d_f-\ad_{\tilde\nu}-\check\omega)U^{-1}(\hat d_f-\ad_{\tilde\nu}-\check\omega)\op^{-1} & -\op^{-1}(\hat d_f-\ad_{\tilde\nu}-\check\omega)U^{-1}\\
-U^{-1}(\hat d_f-\ad_{\tilde\nu}-\check\omega)\op^{-1} & U^{-1}
\end{array}
\right), \nonumber
\end{align}
where $U=\bar P^T(\hat d_f-\ad_{\tilde\nu}-\check\omega)\bar P-\bar P^T(\hat d_f-\ad_{\tilde\nu}-\check\omega)\op^{-1}(\hat d_f-\ad_{\tilde\nu}-\check\omega)\bar P$. 

Note that $\tilde{\mathfrak g}=\{x\in\check{\mathfrak g}\,|\,\Str(x\lambda)=0\,,\,\forall\lambda\in\bar{V}^*\}$ is indeed a subalgebra since for $x,y\in\tilde{\mathfrak g}$ we have $\Str([x,y]\lambda)=-\Str(x\ad_\lambda y)=0$ as a consequence of (\ref{eq:GlobalShift}). In fact for $x,y\in\check{\mathfrak g}$ we have in the same way $[x,y]\in\tilde{\mathfrak g}$. This means in particular that if $\bar{V}$ closes into a subalgebra it must be abelian. Clearly $\check\omega$ reduces to a 2-cocycle $\tilde\omega=\tilde P^T\check\omega\tilde P$ on $\tilde{\mathfrak g}$.

After some algebra and dropping a total derivative $d\nu d\bar\rho$-term, the dualised action becomes 
\begin{align}
-T\int & d^2\sigma\Str\Big(
(J_++\partial_+\bar\rho)\hat d_f(J_-+\partial_-\bar\rho)
+(\partial_+\tilde\nu-\hat d_f^TJ_+)\op^{-1}(\partial_-\tilde\nu+\hat d_fJ_-)
\nonumber\\
&{}
+(\partial_+\tilde\nu-\hat d_f^TJ_+)\op^{-1}(\hat d_f-\ad_{\tilde\nu}-\check\omega)\partial_-\bar\rho
-\partial_+\bar\rho(\hat d_f-\ad_{\tilde\nu}-\check\omega)\op^{-1}(\partial_-\tilde\nu+\hat d_fJ_-)
\nonumber\\
&{}
-\partial_+\bar\rho(\hat d_f-\ad_{\tilde\nu}-\check\omega)\op^{-1}(\hat d_f-\ad_{\tilde\nu}-\check\omega)\partial_-\bar\rho
-\partial_+\bar\rho(\ad_{\tilde\nu}+\check\omega)\partial_-\bar\rho
\Big)\,.
%
\end{align}
As expected $\bar\nu=\check\nu-\tilde\nu$ has dropped out, since we have dualised the corresponding directions. Finally $\bar\rho$ can be removed by the field redefinition
\begin{equation}
f\rightarrow \bar hf\,,\quad\tilde\nu\rightarrow\tilde P^T\left(\Ad_{\bar h}\nu+\frac{1-\Ad_{\bar h}}{\ad_{\bar\rho}}\check\omega\bar\rho\right)\,,\qquad \bar h=e^{-\bar\rho}\,,
\end{equation}
which resembles a $\tilde G$ gauge transformation except for the fact that $\bar h\notin\tilde{G}$. 
To check that we match with the DTD action in~\eqref{eq:S-DTD} we use the fact that under the above redefinition $\op\rightarrow\check P^T\Ad_{\bar h}\op\Ad_{\bar h}^{-1}\check P$ which follows from\footnote{These are proved using~\eqref{eq:admu},~\eqref{eq:dmu} and $\tilde P\Ad_{\bar h}\check P=\Ad_{\bar h}\tilde P$, the last being a consequence of $[x,y]\in\tilde{\mathfrak g}$ for any $x,y\in\check{\mathfrak g}$.}
\begin{equation}
\begin{aligned}
\check P^T\ad_{\tilde\nu}\check P&\rightarrow
\check P^T\Ad_{\bar h}\check P^T\ad_\nu\check P\Ad_{\bar h}^{-1}\check P
+\check P^T\Ad_{\bar h}\check\omega\Ad_{\bar h}^{-1}\check P
-\check\omega\,,
\\
d\tilde\nu&\rightarrow\check P^T\Ad_{\bar h}(d\nu-\ad_\nu(\bar h^{-1}d\bar h)-\check\omega(\bar h^{-1}d\bar h))\,.
\end{aligned}
\end{equation}
The calculations are simple when $\bar{V}$ is a (abelian) subalgebra since in that case $\bar h^{-1}d\bar h=-\Ad_{\bar h}^{-1}d\bar\rho$ and the last $d\bar\rho d\bar\rho$ term vanishes up to a total derivative. 
When $\bar{V}$ is not a subalgebra it is clear that it must still work since these are abelian isometries and we can just T-dualise one at a time. It is nevertheless instructive to show this explicitly. To do this we use the fact that $\bar h^{-1}d\bar h+\Ad_{\bar h}^{-1}d\bar\rho$ is in $\tilde{\mathfrak g}$ since it involves commutators of elements from $\bar{V}$. This simplifies the left-over terms to $\int d\sigma^2\epsilon^{ij}\Str(\bar h^{-1}\partial_i\bar h\ \check\omega(\bar h^{-1}\partial_j\bar h))$ 
which indeed is a total derivative term and can be dropped.
As anticipated, we get that T-dualising on the shift isometries is equivalent to shrinking $\tilde{\mathfrak g}$ by removing the generators in $\bar V$.

\paragraph{Dualising type 1 and 2 isometries}
We have seen that dualising on the isometries outside of $\tilde{\mathfrak g}$ has the effect of adding the corresponding generators to $\tilde{\mathfrak g}$. Similarly dualising on isometries inside $\tilde{\mathfrak g}$ effectively removes the corresponding generators. The natural conjecture is then that dualising on both types of isometries at the same time again just adds/removes the generators outside/inside $\tilde{\mathfrak g}$ to give the $\tilde{\mathfrak g}$ of the resulting model. 

To be more specific, start from a DTD model with a cocycle on the subalgebra\footnote{Also here we prefer to change notation and call $\check{\mathfrak g}$ the original subalgebra, so that $\tilde{\mathfrak g}$ will be used for the algebra obtained after applying NATD.} $\check{\alg{g}}$ and imagine the most general NATD of this DTD model where we dualise isometries $t_i\notin\check{\mathfrak g}$ of type 1 as in (\ref{eq:GlobalG}) and $\lambda_I\in\check{\mathfrak g}^*$ of type 2 as in (\ref{eq:GlobalShift}). Our conjecture is that this results in a new DTD model where now 
\begin{equation}
\tilde{\mathfrak g}=\{x=\check y+a_it_i\,,\,\check y\in\check{\mathfrak g}\,|\,\Str(\lambda_I\check y)=0\,,\,\forall\lambda_I\,\,\,\mbox{such that}\,\,\,\Str(\lambda_I[t_i,t_j])=0\,\,,\forall t_i,t_j\}\,.
\label{eq:dual-gtilde}
\end{equation}
In other words, $\tilde{\mathfrak g}$ is obtained by adding to $\check{\mathfrak g}$ all generators $t_i$ and by removing all elements which are dual to $\lambda_I$, except when these are generated in commutators $[t_i,t_j]$. In fact, we want the last condition on $\lambda_I$ because the commutator of two isometries of type 1 can generate an isometry of type 2, and if we are adding the $t_i$ we want to make sure that they close into an algebra. Here we will not work out explicitly the transformation of the action under this NATD {since this is quite involved},  we will rather just check that this expectation makes sense and such a DTD model is well-defined. 

To start, we must assume that the isometries on which we dualise form a subalgebra of the isometry algebra. This implies the conditions
\begin{equation}
[t_i,t_j]=c_{ij}{}^kt_k+\check c_{ij}{}^{K'}\check t_{K'}\,,\qquad\check\omega(\check t_{I'})=\delta_{I'}^I\lambda_I\,,\qquad\check P^T\ad_{t_i}\lambda_I=c_{iI}{}^J\lambda_J\,,
\label{eq:iso-subalg}
\end{equation}
with some coefficients $c_{ij}{}^k$, $\check c_{ij}{}^k$ and $c_{iI}{}^J$. {The generators $\check t_{K'}\in\check{\mathfrak g}$ appear because, as already mentioned, the commutators of two $t_i$ can generate an element in $\check{\mathfrak g}$. These must still satisfy the second condition in (\ref{eq:GlobalG}) which translates to the second condition above.} The first consistency check is to show that $\tilde{\mathfrak g}$ defined above indeed forms a subalgebra of $\mathfrak g$ so that the corresponding DTD model can be defined. Commuting two elements of $\tilde{\mathfrak g}$ we get
\begin{equation}
[\check y+a_it_i,\check z+b_jt_j]=
[\check y,\check z]
-b_i\ad_{t_i}\check y
+a_i\ad_{t_i}\check z
+a_ib_j[t_i,t_j]\,.
\end{equation}
The isometry conditions in (\ref{eq:GlobalG}) indeed imply that the second and third term are in $\check{\mathfrak g}$. Taking the supertrace with $\lambda_I$ satisfying $\Str(\lambda_I[t_i,t_j])=0$ we get
\begin{equation}
\Str([\check y,\check z]\lambda_I)
+b_ic_{iI}{}^J\Str(\check y\lambda_J)
-a_ic_{iI}{}^J\Str(\check z\lambda_J)
=
-\Str(\check y\ad_{\lambda_I}\check z)
=0\,,
\end{equation}
where we used the conditions (\ref{eq:iso-subalg}) and the fact that $\check y,\check z\in\tilde{\mathfrak g}$ and, in the last step, the isometry condition (\ref{eq:GlobalShift}) for $\lambda_I$. This proves that indeed $\tilde{\mathfrak g}$ in (\ref{eq:dual-gtilde}) defines a subalgebra of $\mathfrak g$. 
To define a 2-cocycle on $\tilde{\mathfrak g}$ we take $\omega=\tilde P^T\check\omega\tilde P$---we could also add an additional deformation in the $t_i$ directions but we will not do so here--- and we find
\begin{align}
\omega[\check y+a_it_i,\check z+b_jt_j]
=&\,
\tilde P^T
\Big(
[\check\omega\check y,\check z+b_it_i]
+[\check y+a_it_i,\check\omega\check z]
+a_ib_j\check\omega[t_i,t_j]
\Big)
\nonumber\\
=&\,
\tilde P^T[\omega\check y,\check z+b_it_i]
+\tilde P^T[\check y+a_it_i,\omega\check z]
+a_ib_j\tilde P^T\check\omega[t_i,t_j]\,,
\end{align}
where we used the cocycle condition for $\check{\omega}$, the fact that $\ad_{t_i}$ commutes with $\check\omega$ (\ref{eq:GlobalG}), and in the last step we used (\ref{eq:PPT-rel}). The first two terms are precisely what we want, it remains to show that the last one vanishes. By the conditions (\ref{eq:iso-subalg}) this term is proportional to a combination of $\lambda_I$ and therefore the $\tilde P^T$ projection means that this term vanishes unless $\Str([t_k,t_l]\check\omega[t_i,t_j])\neq0$ for some $k,l$. However
\begin{align}
\Str([t_k,t_l]\check\omega[t_i,t_j])
=&\,
\tfrac12\Str(\check\omega[[t_i,t_j],[t_k,t_l]])
=
\tfrac12\Str(\check P^T[\check\omega[t_i,t_j],[t_k,t_l]])
+\tfrac12\Str(\check P^T[[t_i,t_j],\check\omega[t_k,t_l]])
\nonumber\\
=&\,
\tfrac12\check c_{ij}{}^I\Str(\check P^T\ad_{\lambda_I}[t_k,t_l])
-\tfrac12\check c_{kl}{}^I\Str(\check P^T\ad_{\lambda_I}[t_i,t_j])
=0\,,
\end{align}
where we used the cocycle condition and the isometry condition in (\ref{eq:GlobalShift}). Therefore $\omega$ is indeed a 2-cocycle on $\tilde{\mathfrak g}$ and the corresponding DTD model is well-defined.

\subsection{DTD models not related to YB models by NATD}\label{sec:not-YB}
Here we want to present an example of a DTD model which is not related to a YB model by NATD.\footnote{{Let us mention that it is possible to find examples where $\omega$---as well as any 2-cocycle in its equivalence class---is non-degenerate on a space which does not close into an algebra. This corrects a statement in the first version of~\cite{Borsato:2016pas}.}}
To argue that this is the case we use two important facts concerning the dualisation of the two types of isometries discussed above.
First, when dualising isometries of type 1, thanks to the condition~\eqref{eq:GlobalG} the original $\check{\mathfrak g}$ will become an ideal of the larger algebra $\tilde{\mathfrak g}$ that is obtained by adding the generators $t_i$, i.e. by applying NATD.
That means that starting from a YB model---or, rather, its corresponding DTD model with non-degenerate $\omega$---NATD on isometries of type 1 will produce a DTD model with a cocycle \emph{non-degenerate on an ideal} of $\tilde{\mathfrak g}$. When we include also isometries of type 2 it remains true that what is left of $\check{\mathfrak g}$ forms a proper ideal inside $\tilde{\mathfrak g}$, on which, however, $\omega$ does not have to be non-degenerate.
We also remark that, since they are realised as linear shifts, isometries of type 2 are {commuting} and are therefore still present even after applying {abelian} T-duality along them. After the dualisation the corresponding symmetry will be realised as an isometry of type 1.

Consider the following algebra and corresponding 2-cocycle
\begin{equation}
\tilde{\mathfrak g}=\mathrm{span}\{p_1,\,p_2,\,p_3,\,J_{12}\}\,,\qquad\omega=k_3\wedge J_{12}\,,
\label{eq:ex1}
\end{equation}
where we refer to \cite{Borsato:2016ose} for our definitions and conventions on the generators of the conformal algebra {$\mathfrak{so}(2,4)$}.
The above 2-cocycle is defined on a space which is not an ideal of $\tilde{\mathfrak g}$, and it is clear that {adding an exact term to $\omega$ cannot change this}, since the only terms that we could add are $k_1\wedge J_{12}$ and $k_2\wedge J_{12}$. 
According to the above discussion, this rules out the possibility of this example coming from dualising isometries of type 1 of a YB model. {In fact, since there is no proper ideal in $\tilde{\mathfrak g}$ that contains the subspace $\{p_3,J_{12}\}$ where $\omega$ is defined, a combination of isometries of type 1 and type 2 is also ruled out. This leaves only the possibility that this example is generated by T-dualising isometries of type 2 only. If} it were true that it comes from a YB model by dualising isometries of type 2, these should be realised here as isometries of type 1 and we {would be able to dualise them back to find a YB model (in DTD form)}. 
However, in this example the only isometry of type 1 corresponds to $p_0$, and adding $p_0$ to $\tilde{\mathfrak g}$ does not help in making the cocycle non-degenerate on the dualised algebra.
We therefore conclude that the above example is not related to a YB model by NATD,\footnote{It would be interesting to understand whether this or similar examples are related to YB models in other ways, e.g. contractions.} and we refer to section~\ref{sec:ex-not-YB} {for the corresponding supergravity background}.

The above example may be obtained by dropping one of the two terms in $R_{11}$ {in table 2} of \cite{Borsato:2016ose}, and similar examples coming from dropping a term in other rank 4 R-matrices of \cite{Borsato:2016ose} are e.g. 
{
\begin{equation}
\begin{aligned}
&\tilde{\mathfrak g}=\mathrm{span}\{p_1,\,p_2,\,p_3,\,p_0+J_{12}\}\,,\qquad&&\omega=k_3\wedge({k_0+}\, J_{12})\,,&&&\text{ from } R_{10}\,.\\
&\tilde{\mathfrak g}=\mathrm{span}\{p_0,\,p_1,\,p_2,\,J_{12}\}\,,\qquad&&\omega=k_0\wedge J_{12}\,,&&&\text{ from } R_{13}\,.\\
&\tilde{\mathfrak g}=\mathrm{span}\{p_1,\,p_2,\,J_{12},\,J_{03}\}\,,\qquad&&\omega=J_{12}\wedge J_{03}\,,&&&\text{ from } R_{14}\,.
\end{aligned}
\end{equation}
In each case it is easy to see that $\omega$ cannot be defined on an ideal in $\tilde{\mathfrak g}$ even if we add exact terms---in the first case the only terms that we could add are $k_1\wedge({k}_0+J_{12})$ and $k_2\wedge({k}_0+J_{12})$, in the second and third case they are $k_1\wedge J_{12}$ and $k_2\wedge J_{12}$. In the first case the only isometry of type 1 corresponds to $p_0$, while in the second and third there is no isometry of type 1. Note that the second case can be embedded into $\mathfrak{so}(2,3)$ and therefore gives a deformation also of $AdS_4$.
}

\section{Kappa symmetry and Green-Schwarz form}\label{sec:kappa}
As we will show in a moment the action of DTD models is invariant under kappa symmetry variations, and this will allow us to put it into the Green-Schwarz form. To show invariance under kappa symmetry we need to consider the variation of the action under the fields $\nu$ and $f$, as well as the worldsheet metric $\gamma^{ij}$. 
The variation of the action with respect to the fields is computed in~\eqref{eq:delta-fnu-S}.
To define a kappa symmetry variation we should also say how $\delta f$ and $\delta \nu$ are expressed in terms of the kappa symmetry parameters $\tilde\kappa^{(j)}_i$, each of them being a local Grassmann parameter of grading $j$. We define $A_\pm^i\equiv \Ad_f^{-1}(\tilde A_\pm^i+J_\pm^i)$, where subscripts $\pm$ indicate that we act with the worldsheet projectors in~\eqref{eq:ws-proj} and $\tilde A^i_\pm$ is given in~\eqref{eq:solApm}; we take\footnote{We write the kappa symmetry transformation in this way rather than the one in~\cite{Borsato:2016pas} because we want  $P^{(0)}\Ad_f^{-1}\delta_\kappa\nu=0$.}
\be
\hat d^T(f^{-1}\delta_\kappa f)=\Ad_f^{-1}\delta_\kappa \nu=
-\{i\tilde\kappa^{(1)}_i,A_-^{(2)i}\}+\{i\tilde\kappa^{(3)}_i,A_+^{(2)i}\}\,.
\ee
This relation is fixed by noticing that after we impose it the total variation of the action with respect to the fields simplifies considerably, and we find
\be
\begin{aligned}
(\delta_f+\delta_\nu )S&=-\tfrac{T}{2}\int d^2\sigma \ 4\STr\left(
A_-^{(2)i}A_-^{(2)j}[A_{+i}^{(1)},i\tilde\kappa_j^{(1)}]
+A_+^{(2)i}A_+^{(2)j}[A_{-i}^{(3)},i\tilde\kappa_j^{(3)}]
\right)\\
&=-\tfrac{T}{2}\int d^2\sigma \ \tfrac{1}{2}\Big[
\STr\left(A_-^{(2)i}A_-^{(2)j}\right)\STr\left(W[A_{+i}^{(1)},i\tilde\kappa_j^{(1)}]\right)\\
&\qquad\quad\qquad\qquad+\STr\left(A_+^{(2)i}A_+^{(2)j}\right)\STr\left(W[A_{-i}^{(3)},i\tilde\kappa_j^{(3)}]\right)
\Big]\,.
\end{aligned}
\ee
Here we used the property $A_\pm^i B_\pm^j=A_\pm^j B_\pm^i$, which follows from the identity $ P^{ij}_\pm P^{kl}_\pm= P^{il}_\pm P^{kj}_\pm$, as well as the identity
\be
A_\pm^{(2)i}A_\pm^{(2)j} = \tfrac{1}{8}W\STr(A_\pm^{(2)i}A_\pm^{(2)j})+c^{ij}\mathbbm1_8\,,
\ee
where $c^{ij}$ is an expression which is not interesting for this calculation, and $W=\text{diag}(1_4,-1_4)$ is the hypercharge.
The above variation does not vanish but it can be compensated by the contribution coming from varying the worldsheet metric.
In fact, we first notice that the contribution of the terms involving the worldsheet metric to the action may be written as
\be\label{eq:Sgamma}
S_\gamma=-\tfrac{T}{2}\int d^2\sigma \gamma^{ij}\ \STr\left(E^{(2)}_i E^{(2)}_j\right)\,,
\ee
where we have two possible choices for the bosonic vielbein which are related by a local Lorentz transformation, either $E^{(2)}=A_+^{(2)}$ or $E^{(2)}=A_-^{(2)}$, where 
\begin{equation}\label{eq:def-Apm}
A_+=\Ad_f^{-1}(J+\op^{-T}(d\nu-\hat d_f^TJ))\,,\qquad
A_-=\Ad_f^{-1}(J-\op^{-1}(d\nu+\hat d_fJ))\,.
\end{equation}
The subscript on $A_\pm$ is here used only to distinguish the two fields and should not be confused with the $\pm$ used to denote the worldsheet projections; however, we choose this notation since projecting on $A_\pm$ with $ P_\pm^{ij}$ after reintroducing worldsheet indices we obtain in fact the $A_\pm^{i}$ used above.\footnote{A caveat is that the projections of $A_\pm$ in~\eqref{eq:def-Apm} with $P_\mp^{ij}$ do not vanish, while $P_\mp^{ij}A_{\pm j}=0$. We trust that this will not create confusion, since the notation has clear advantages and those projections will never be needed.}
We declare the kappa symmetry variation of the worldsheet metric to be
\be
\delta_\kappa\gamma^{ij}=-\tfrac{1}{2}\left[\STr\left(W[A_{+}^{(1)i},i\tilde\kappa_+^{(1)j}]\right)
+\STr\left(W[A_{-}^{(3)i},i\tilde\kappa_-^{(3)j}]\right)\right]\,,
\ee
so that the total variation of the action under kappa symmetry transformations vanishes $(\delta_f+\delta_\nu +\delta_\gamma ) S=0$.
The kappa symmetry transformations for the fields may be also recast into the form
\begin{equation}\label{eq:deltakappa}
i_{\delta_\kappa z}E^{(2)}=0\,,\qquad
i_{\delta_\kappa z}E^{(1)}=P^{ij}_-\{i\kappa^{(1)}_i,E_j^{(2)}\}\,,\qquad
i_{\delta_\kappa z}E^{(3)}=P^{ij}_+\{i\kappa^{(3)}_i,E_j^{(2)}\}\,,
\end{equation}
where $\kappa^{(1)}=\Ad_h\tilde\kappa^{(1)}$ and $\kappa^{(3)}=\tilde\kappa^{(3)}$ and where we made a choice for the bosonic and fermionic components of the supervielbeins
\begin{equation}\label{eq:E-A}
E^{(2)}=A_+^{(2)}=\Ad_hA_-^{(2)}\,,\qquad E^{(1)}=\Ad_hA_+^{(1)}\,,\qquad E^{(3)}=A_-^{(3)}\,.
\end{equation}
The above transformations are the standard ones for kappa symmetry, and the action also takes the standard Green-Schwarz form
\begin{equation}\label{eq:S-GS}
S=-\tfrac{T}{2}\int d^2\sigma\,\gamma^{ij}\Str(E_i^{(2)}E_j^{(2)})-T\int B\,,
\end{equation}
where the $B$-field is
\begin{equation}\label{eq:B-field}
B=\tfrac14\Str(J\wedge\hat d_fJ+(d\nu-\hat d_f^TJ)\wedge\op^{-1}(d\nu+\hat d_fJ))\,.
\end{equation}

As already noticed, $A_+^{(2)}$ and $A_-^{(2)}$ are related by a local Lorentz transformation, $A_+^{(2)}=\Ad_hA_-^{(2)}$ for some $h\in G^{(0)}$. For later convenience we can also relate other components of $A_+$ and $A_-$ as follows\footnote{As a consequence of this we have for example $A_+^{(3)}=E^{(3)}-P^{(3)}ME^{(2)}$.}
\begin{align}\label{eq:M}
&A_-=MA_+\,,\qquad P^{(2)}M=\Ad_h^{-1}P^{(2)}
\,,
\\
&M=\Ad_f^{-1}[1-\tilde P-\op^{-1}\op^T-4\op^{-1}\Ad_fP^{(2)}\Ad_f^{-1}(1-\tilde P)]\Ad_f=
1-4\Ad_f^{-1}\op^{-1}\Ad_fP^{(2)}\,,
\nonumber
\end{align}
while $M^{-1}$ is given by the same expression as $M$ but with $\op$ replaced by its transpose $\op^T=\tilde P^T(\hat d_f^T+\ad_\nu+\zeta\omega)\tilde P$. From this we can derive the useful relation
\begin{equation}
M^{-1}-1=-(M-1)\Ad_h\,.
\end{equation}

\section{Target space superfields}\label{sec:target}
In this section we will derive the form of the target space supergravity superfields for the DTD model. The calculations are very similar to the ones performed in \cite{Borsato:2016ose} for the $\eta$-model and $\lambda$-model.
Once the action and kappa symmetry transformations are written in Green-Schwarz form as in~\eqref{eq:S-GS} and~\eqref{eq:deltakappa}, the easiest way to extract the background fields is by computing the torsion $T^a=dE^a+E^b\wedge\Omega_b{}^a$ and $T^\alpha=dE^\alpha-\frac14(\Gamma_{ab}E)^\alpha\wedge\Omega^{ab}$ where $\Omega^{ab}$ is the spin connection superfield. It was shown in \cite{Wulff:2016tju} that the constraints on the torsion implied by kappa symmetry take the form\footnote{This is valid only for a suitable choice of the spin connection, which can however be extracted from the same equations. We have dropped the $\wedge$'s for readability.}
\begin{align}
T^a=-\tfrac{i}{2}E\gamma^aE\,,\quad
T^{\alpha I}
=&\,
\tfrac12E^{\alpha I}\,E\chi
+\tfrac12(\sigma^3E)^{\alpha I}\,E\sigma^3\chi
-\tfrac14E\gamma_aE\,(\gamma^a\chi)^{\alpha I}
-\tfrac14E\gamma_a\sigma^3E\,(\gamma^a\sigma^3\chi)^{\alpha I}
\nonumber\\
&\qquad{}
-\tfrac18E^a\,(E\sigma^3\gamma^{bc})^{\alpha I} H_{abc}
-\tfrac18E^a\,(E\gamma_a\mathcal S)^{\alpha I}
+\tfrac12E^bE^a\,\psi^{\alpha I}_{ab}\,,
\label{eq:torsion}
\end{align}
for the type IIB case.\footnote{Essentially identical expressions hold for type IIA, cf. \cite{Wulff:2013kga}.} The target space superfields contained here are the dilatino superfields $\chi_{\alpha I}$, the gravitino field strengths $\psi_{ab}^{\alpha I}$, where $I=1,2$ denotes the two Majorana-Weyl spinors of type IIB, as well as the NSNS three-form field strength $H=dB$ and ``RR field strengths'' encoded in the anti-symmetric $32\times32$ bispinor
\begin{equation}
\mathcal S=-i\sigma^2\gamma^a\mathcal F_a-\tfrac{1}{3!}\sigma^1\gamma^{abc}\mathcal F_{abc}-\tfrac{1}{2\cdot5!}i\sigma^2\gamma^{abcde}\mathcal F_{abcde}\,.
\label{eq:calS}
\end{equation}
Kappa symmetry implies that the target space is generically only a solution of the \emph{generalised} type II supergravity equations defined in \cite{Wulff:2016tju} and first written down, for the bosonic sector, in~\cite{Arutyunov:2015mqj}. However, when the (Killing) vector
\begin{equation}
K^a=-\tfrac{i}{16}(\gamma^a\sigma^3)^{\alpha I\beta J}\nabla_{\alpha I}\chi_{\beta J}
\label{eq:Ka}
\end{equation}
vanishes one gets a solution of \emph{standard} type II supergravity, and a one-loop Weyl invariant string sigma model. In that case there exists a dilaton superfield $\phi$ such that $\chi_{\alpha I}=\nabla_{\alpha I}\phi$ and the RR field strengths are defined in terms of potentials in the standard way $\mathcal F=e^\phi dC+\cdots$ \cite{Tseytlin:1996hs,Wulff:2013kga}.

Given that the supervielbeins for the DTD model are defined in terms of $A_\pm$ as in (\ref{eq:E-A}) we need to compute the exterior derivative of $A_\pm$ defined in (\ref{eq:def-Apm}) to find the torsion. With a bit of work one finds the deformed ``Maurer-Cartan'' equations\footnote{We use anti-commutators rather than commutators because the objects that appear are one-forms, and therefore naturally anti-commute.}
\begin{align}
dA_+=&\,
\tfrac12\{A_+,A_+\}
-\tfrac12\Ad_f^{-1}\op^{-T}\Ad_f\big(\hat d^T\{A_+,A_+\}-2\{A_+,\hat d^TA_+\}\big)\,,
\label{eq:dA+}
\\
dA_-=&\,
\tfrac12\{A_-,A_-\}
-\tfrac12\Ad_f^{-1}\op^{-1}\Ad_f\big(\hat d\{A_-,A_-\}-2\{A_-,\hat dA_-\}\big)\,,
\label{eq:dA-}
\end{align}
where we have used the identity~\eqref{eq:PPT-rel} and the fact that, due to the Jacobi identity and the 2-cocycle condition~\eqref{eq:cocycle2}, both $\ad_\nu$ and $\omega$ effectively act as derivations on the Lie bracket. 
Projecting the first equation with $P^{(2)}$ and using~\eqref{eq:E-A} and (\ref{eq:M}) we get
\begin{align}
dE^{(2)}=&
\{A_+^{(0)},E^{(2)}\}
+\tfrac12\{E^{(1)},E^{(1)}\}
+\tfrac12\{E^{(3)},E^{(3)}\}
-\{E^{(3)},P^{(3)}ME^{(2)}\}
-P^{(2)}M^T\{E^{(2)},E^{(3)}\}
\nonumber\\
&{}
+\tfrac12\{P^{(3)}ME^{(2)},P^{(3)}ME^{(2)}\}
+P^{(2)}M^T\{E^{(2)},P^{(3)}ME^{(2)}\}
-\tfrac12P^{(2)}M^T\{E^{(2)},E^{(2)}\}\,.
%
\end{align}
Using $A_+^{(0)}=\frac12A_+^{ab}J_{ab}$, $E^{(2)}=E^aP_a$ etc. and the algebra in appendix A of \cite{Borsato:2016ose} this gives the form for the bosonic torsion $T^a$ in~\eqref{eq:torsion} provided that we identify the spin connection with\footnote{The components of $M$ are defined as $MT_A=T_BM^B{}_A$.}
\begin{equation}
\Omega_{ab}
=
(A_+)_{ab}
+2i(E^2\gamma_{[a})_{\beta}M^{\beta2}{}_{b]}
+\tfrac{3i}{2}E^cM^{\alpha2}{}_{[a}(\gamma_b)_{\alpha\beta}M^{\beta2}{}_{c]}
+\tfrac12E^c(M_{ab,c}-2M_{c[a,b]})\,.
\end{equation}
In a similar way, using (\ref{eq:E-A}) and (\ref{eq:dA-}) we find that
\begin{align}
dE^{(3)}=&
\{A_+^{(0)},E^{(3)}\}
+\{P^{(0)}ME^{(2)},E^{(3)}\}
+\Ad_h^{-1}\{E^{(1)}+P^{(1)}\Ad_hME^{(2)},E^{(2)}\}
+\tfrac12P^{(3)}M\{E^{(3)},E^{(3)}\}
\nonumber\\
&{}
+2P^{(3)}\Ad_f^{-1}\op^{-1}\Ad_f\Big(
2\Ad_h^{-1}\{E^{(1)}+P^{(1)}\Ad_hME^{(2)},E^{(2)}\}
+\Ad_h^{-1}\{E^{(2)},E^{(2)}\}
\Big),
\end{align}
which leads to the torsion $T^{\alpha 2}$ taking the form in (\ref{eq:torsion}) with the background fields given by\footnote{These expressions have obvious close analogies with the ones found for the  $\eta$-model in \cite{Borsato:2016ose}.}
\begin{align}
H_{abc}=&3M_{[ab,c]}-3iM^{\alpha2}{}_{[a}(\gamma_b)_{\alpha\beta}M^{\beta2}{}_{c]}
\,,\quad
\mathcal S^{\alpha1\beta2}=
-8i[\Ad_h(1+4\Ad_f^{-1}\op^{-T}\Ad_f)]^{\alpha1}{}_{\gamma1}\widehat{\mathcal K}^{\gamma1\beta2},
\label{eq:HandS}
\\
\chi^2_\alpha=&-\tfrac{i}{2}\gamma^a_{\alpha\beta}M^{\beta2}{}_a
\,,\quad
\psi_{ab}^{\alpha2}
=
2[\Ad_f^{-1}\op^{-1}\Ad_f\Ad_h^{-1}]^{\alpha2}{}_{cd}\widehat{\mathcal K}_{ab}{}^{cd}
+\tfrac14[\Ad_hM]^{\beta1}{}_{[a}(\gamma_{b]}\mathcal S^{12})_\beta{}^\alpha\,.
\nonumber
\end{align}
Here $\widehat{\mathcal K}^{AB}$ denotes the inverse of the metric defined by the supertrace $\Str(T_AT_B)=\mathcal K_{AB}$, see appendix A of \cite{Borsato:2016ose} for more details on our conventions. 

Since the DTD model contains NATD as a special case we obtain as a by-product the transformation rules for RR fields under NATD---starting from a supercoset model. As a check we can compare this to the formula conjectured in \cite{Sfetsos:2010uq} based on analogy to the abelian case \cite{Hassan:1999bv}---consistency of that formula was checked in some particular cases {also} in \cite{Hoare:2016wca}. Setting $\zeta=0$, which removes the deformation, and restricting to a bosonic $\tilde{\mathfrak g}$, so that $\tilde P=\tilde P(P^{(0)}+P^{(2)})=(P^{(0)}+P^{(2)})\tilde P$, we find\footnote{Note that $(P^{(0)}+P^{(2)})\Ad_fP^{(1)}=0+$fermions.}
\begin{equation}
\mathcal S^{\alpha1\beta2}=
-8i[\Ad_h|_{\theta=0}]^{\alpha1}{}_{\gamma1}\widehat{\mathcal K}^{\gamma1\beta2}
+\mbox{fermions}\,,
\end{equation}
which agrees with the transformations conjectured in \cite{Sfetsos:2010uq}. Note that our result generalises this to the case where also fermionic T-dualities are involved.

Finally we must compute $T^{\alpha1}$ to extract the other dilatino superfield $\chi^1$. We find
\begin{align}
dE^{(1)}
=&
\{\Ad_hA_+^{(0)}-dhh^{-1},E^{(1)}\}
+\Ad_h\{E^{(2)},E^{(3)}-P^{(3)}ME^{(2)}\}
+\tfrac12P^{(1)}\Ad_hM^{-1}\Ad_h^{-1}\{E^{(1)},E^{(1)}\}
\nonumber\\
&{}
+2P^{(1)}\Ad_h\Ad_f^{-1}\op^{-T}\Ad_f\Big(
2\{E^{(2)},E^{(3)}-P^{(3)}ME^{(2)}\}
+\{E^{(2)},E^{(2)}\}
\Big)\,.
\end{align}
Taking the exterior derivative of the equation $A_+^{(2)}=\Ad_hA_-^{(2)}$, cf. (\ref{eq:M}), we find the relation
\begin{equation}
[\Ad_hA_+^{(0)}-dhh^{-1}]_{ab}
=
\Omega_{ab}
-\tfrac12E^cH_{abc}
+2i(E^1\gamma_{[a})_\alpha[\Ad_hM]^{\alpha1}{}_{b]},
\end{equation}
which can be used to show that the torsion again takes the form in (\ref{eq:torsion}), where the remaining components of the background fields are\footnote{Just as in \cite{Borsato:2016ose}, one finds a superficially different expression for $H_{abc}$ namely
$$
H_{abc}=3[\Ad_hM]_{[ab,c]}+3i[\Ad_hM]^{\alpha1}{}_{[a}(\gamma_b)_{\alpha\beta}[\Ad_hM]^{\beta1}{}_{c]}\,.
$$
However consistency requires this to be the same as the expression in (\ref{eq:HandS}) and this can also be verified explicitly similarly to \cite{Borsato:2016ose}.
}
\begin{equation}
\chi^1_\alpha=\tfrac{i}{2}(\gamma^a)_{\alpha\beta}[\Ad_hM]^{\beta1}{}_a\,,\quad
\psi^{\alpha1}_{ab}
=
2[\Ad_h\Ad_f^{-1}\op^{-T}\Ad_f]^{\alpha1}{}_{cd}\widehat{\mathcal K}_{ab}{}^{cd}
-\tfrac14(\mathcal S^{12}\gamma_{[a})^\alpha{}_\beta M^{\beta2}{}_{b]}\,.
\label{eq:chi1}
\end{equation}
It remains only to analyse the question of when this is a solution to the standard or the generalised type II supergravity equations, in other words to identify the conditions under which $K^a$ defined in (\ref{eq:Ka}) vanishes. We do this in the next subsection.

\subsection{Supergravity condition and dilaton}
By analogy with the calculations performed in \cite{Borsato:2016ose} there is a natural candidate for the dilaton superfield for the DTD model namely\footnote{The prime on the superdeterminant denotes the fact that we must restrict to the subspace where $\op$ is defined, i.e. the subalgebra $\tilde{\mathfrak g}$.}
\begin{equation}
e^{-2\phi}=\mathrm{sdet}'\op\,.
\end{equation}
We will now show that this guess is indeed correct by verifying that its spinor derivatives reproduces the dilatini found above.
 Using the formula for the supertrace $\Str\mathcal M=\widehat{\mathcal K}^{AB}\Str(T_A\mathcal MT_B)$ we find
\begin{align}
d\phi
=&
-\tfrac12\Str(d\op\op^{-1})
=
-\tfrac12\widehat{\mathcal K}^{AB}\Str\big\{([J,\hat d^T_fT_A]-\hat d^T_f[J,T_A]+[d\nu,T_A])\op^{-1}T_B\big\}
\nonumber\\
=&
-\tfrac12\widehat{\mathcal K}^{AB}\Str\big\{\big(
[J,\hat d^T_fT_A]
-\hat d^T_f[J,T_A]
+[\Ad_f\hat d^TA_+,T_A]
+[(\ad_\nu+\zeta\omega)(\Ad_fA_+-J),T_A]
\big)\op^{-1}T_B
\big\}
\nonumber\\
=&
\tfrac12\widehat{\mathcal K}^{AB}\Str\big\{
T_A\big(
\hat d[A_+,\Ad_f^{-1}\op^{-1}\Ad_fT_B]
+[\hat d^TA_+,\Ad_f^{-1}\op^{-1}\Ad_fT_B]
-[A_+,\hat d\Ad_f^{-1}\op^{-1}\Ad_fT_B]
\big)
\big\}
\nonumber\\
&{}
+\widehat{\mathcal K}^{AB}\Str\big\{[(\Ad_fA_+-J),T_A]\tilde PT_B\big\}\,.
\label{eq:dphi}
\end{align}
If the last term vanishes, then using (\ref{eq:E-A}), (\ref{eq:chi1}), (\ref{eq:HandS}) and (\ref{eq:M}) one may check that the $E^{(1,3)}$-terms are indeed equal to
\begin{equation}
E^{\alpha1}\chi^1_\alpha+E^{\alpha2}\chi^2_\alpha\,.
\end{equation}
Therefore $\chi_{\alpha I}=\nabla_{\alpha I}\phi$ which implies that $K^a$ in (\ref{eq:Ka}) vanishes and we have a solution to standard type II supergravity.
Since $(\Ad_fA_+-J)\in\tilde{\mathfrak g}$ can be regarded as an arbitrary element of the Lie algebra, the vanishing of the last term in (\ref{eq:dphi}) is equivalent to $f_{AB}{}^A=0$ for the structure constants of $\tilde{\mathfrak{g}}$, i.e. $\tilde{\mathfrak{g}}$ must be unimodular. This condition is therefore sufficient to get a standard supergravity solution. Following a calculation similar to the one done in \cite{Borsato:2016ose}, computing $K^a$ in (\ref{eq:Ka}) and requiring it to vanish one finds that this condition is also necessary.\footnote{{In very special cases it is possible for $K^a$ to decouple from the remaining generalized supergravity equations. One then obtains a background solving both the generalised and standard supergravity equations depending on if $K^a$ is included or not. One such example is the pp-wave solution discussed in Appendix B of \cite{Hoare:2016hwh}.} We thank B. Hoare and S. van Tongeren for pointing this out.}

Our results imply that the DTD model gives a one-loop Weyl invariant string sigma model precisely\footnote{{This is modulo possible subtleties with the special cases mentioned in the previous footnote. One should also note that this condition is true provided one only allows a \emph{local} (Fradkin-Tseytlin) counter-term. If one relaxes this condition one can find a \emph{non-local} counter-term also when $K^a$ is non-zero, since solutions of the generalised supergravity equations are formally T-dual to solutions of the standard ones; see also \cite{Sakamoto:2017wor}. This being said, cases where $K^a$ is null may be subtle and deserve further study.}} when the subalgebra $\tilde{\mathfrak{g}}$ is unimodular. This is in fact the same condition that was found  long  ago for NATD on bosonic sigma models by path integral considerations \cite{Alvarez:1994np,Elitzur:1994ri}.  Since the DTD model includes NATD as a special case, the analysis here coupled with the results of \cite{Arutyunov:2015mqj,Wulff:2016tju}, gives an alternative derivation of the Weyl anomaly for NATD of supercosets. 

A nice fact is that we do not have to impose extra conditions on the cocycle $\omega$ used to construct the deformation. When $\omega$ is non-degenerate unimodularity of $\tilde{\mathfrak{g}}$ is equivalent to unimodularity of $R=\omega^{-1}$ as defined in \cite{Borsato:2016ose}, see the discussion there; this is consistent with the fact that the YB models are a special case of the DTD models.



\section{Some explicit examples}\label{sec:ex}
Here we would like to collect some formulas that are useful when deriving the explicit background for a given DTD model, and then work out {two} examples in detail.
We denote the generators of $\tilde{\alg{g}}\subset \alg g$ by $ T_i$, $i=1,\ldots,N=\text{dim}(\tilde{\alg{g}})$, and those of the dual $\tilde{\alg{g}}^*$ by $ T^i$. They satisfy $\STr(T^iT_j)=\delta^i_j$. The action of the projectors on a generic element  $x\in\alg g$ may be written as
\be
\tilde P(x)=\STr(T^ix)T_i,\qquad
\tilde P^T(x)=\STr(T_ix)T^i,
\ee
where summation of repeated indices is assumed. Given a cocycle $\omega=\tfrac12\omega_{ij}T^i\wedge T^j$ with $\omega_{ji}=-\omega_{ij}$, its action on an element of the algebra is
\be
\omega(x) = \omega_{ij}T^i\STr(T^jx),
\ee
and it must satisfy the cocycle condition, which may be written as
\be
\STr\Big(T_k(\omega[{T_i},T_j]-[{T_i},\omega T_j]+[{T_j},\omega T_i])\Big)=0,
\qquad\quad \forall T_i,T_j,T_k\in\tilde{\alg{g}}.
\ee
With the above definitions one may easily construct the operator $\op:\tilde{\alg{g}}\to\tilde{\alg{g}}^*$ defined in (\ref{eq:def-op}), that can be encoded in an explicit $N\times N$ matrix 
\be
\tilde O_{ij}=\STr(\op(T_i)T_j),
\ee
so that $\op(T_i)=\tilde O_{ij}T^j$.
The matrix $\tilde O$ can be inverted with standard methods and used to construct the action of the inverse operator as $ \op^{-1}(x)=\STr(xT_i)(\tilde O^{-1})^{ij}T_j$, so that on the basis generators $\op^{-1}(T^i)=(\tilde O^{-1})^{ij}T_j$.
Obviously, when choosing a parametrisation for the group element $f$, one should make sure that the corresponding degrees of freedom cannot be gauged away by applying the local transformations discussed in section~\ref{sec:DTD}.

To obtain the background fields we use the results of section~\ref{sec:target}.
The metric reads as $ds^2=\eta^{ab}E_aE_b$, where the components of the bosonic supervielbein are obtained by $E_a=\STr(A_+ P_a)$, and the $B$-field is given by equation~\eqref{eq:B-field}.
From the superdeterminant of the matrix $\tilde O$ it is also straightforward to compute the (exponential of the) dilaton $e^\phi=(\mathrm{sdet}\,\tilde O)^{-\frac{1}{2}}$.
In order to determine the RR fields one first identifies the components of the matrix $M_{ab}=\STr((M{P}_a){P}_b)$ and then one constructs the local Lorentz transformation on spinorial indices
\begin{equation}
{(\Ad_h)^{\beta}}_{\alpha} = \exp[- \tfrac{1}{4} (\log M)_{ab} \Gamma^{ab} ]{^{\beta}}_{\alpha}\,,
\end{equation}
so that $\Ad_h\Gamma_a\Ad_h^{-1} =M_a^{\ b}\Gamma_b$, where $\Gamma_a$ are $32\times 32$ Gamma-matrices\footnote{Alternatively one can use the $16\times16$ gamma matrices used in the previous section.}. From (\ref{eq:calS}) and (\ref{eq:HandS}) one finds that the expression for RR fields is obtained by solving the equation
\begin{equation}
(\Gamma^a F_a + \tfrac{1}{3!} \Gamma^{abc} F_{abc} + \tfrac{1}{2\cdot 5!} \Gamma^{abcde} F_{abcde})\Pi = 
e^{-\phi} \ [\Ad_h(1+4\Ad_f^{-1}\op^{-T}\Ad_f)](4 \Gamma_{01234})\Pi,
\end{equation}
where $\Pi = \tfrac{1}{2} (1-\Gamma_{11})$ is a projector\footnote{With these conventions the self-duality for the 5-form is $F^{(5)}=*F^{(5)}$.} and $(-4 \Gamma_{01234})\Pi$ corresponds to the 5-form flux of AdS$_5\times$S$^5$.
In order to find the component $F_{a_1\ldots a_{2m+1}}$ it is then enough to multiply the above equation by $\Gamma_{a_1\ldots a_{2m+1}}$ and take the trace. As already explained, when the subalgebra $\tilde{\alg{g}}$ is bosonic the above result simplifies considerably, and only $\Ad_h$ remains inside square brackets.
After obtaining the components in tangent indices we translate them into form language  using
$F^{(2m+1)}=\frac{1}{(2m+1)!}E^{a_{2m+1}}\wedge\ldots\wedge E^{a_1} F_{a_1\ldots a_{2m+1}}$.

\subsection{A TsT example}
{First we will work out a simple example where we dualise} a two-dimensional abelian subalgebra of the isometry of the sphere $\alg{so}(6)$, so that the deformation is equivalent to doing a TsT there~\cite{Lunin:2005jy,Frolov:2005ty,Frolov:2005dj}. This example was worked out already in~\cite{Hoare:2016wsk} for the NSNS sector, and the RR fields were taken into account in~\cite{Hoare:2016wca} by following the T-duality rules of~\cite{Sfetsos:2010uq}.
Here we will use the matrix realisation of the $\alg{psu}(2,2|4)$ superalgebra used in~\cite{Borsato:2016ose}, see also~\cite{Arutyunov:2015qva}.
We take $\tilde{\alg{g}}$ to be the abelian algebra spanned by two Cartans of $\alg{so}(6)$, $T_1\equiv J_{68},T_2\equiv J_{79}$, and for the dual generators we may just take $T^1= J_{68},T^2= J_{79}$.
We parametrise the bosonic fields as\footnote{The group elements parametrised by $\varphi$, $\xi$ and $r$ coincide with those in (A.1) of~\cite{Arutyunov:2013ega}.}
\be
\nu = \tilde \varphi_iT^i,\qquad\qquad
f=f_{\alg{a}} \cdot \exp(\varphi P_5)\exp(-\xi J_{89})\exp(-\arcsin rP_9),
\ee
where $f_{\alg{a}}$ is a coset group element parametrised by fields in AdS$_5$.
We take  $\omega=T^1\wedge T^2$ which obviously satisfies the cocycle condition. The matrix corresponding to $\op$ is very simple
\be
\tilde O_{ij}=\left(
\begin{array}{cc}
 2 r^2 \sin ^2\xi  & \zeta  \\
 -\zeta  & 2 r^2 \cos ^2\xi  \\
\end{array}
\right),
\ee
and it is easily inverted. Following the above discussion we immediately find the fields of the NSNS sector
\be
\begin{aligned}
&ds^2=ds_{\alg{a}}^2+
\frac{r^2 }{\zeta ^2+r^4 \sin ^2(2 \xi )}(\cos ^2\xi\, d\tilde\varphi_1^2+\sin ^2\xi\, d\tilde\varphi_2^2)
+(1-r^2)d\varphi^2+r^2d\xi^2+\frac{dr^2}{1-r^2}\,,
\\
&e^\phi=(\zeta ^2+r^4 \sin ^2(2 \xi ))^{-\frac{1}{2}}\,,\qquad B=\frac{\zeta}{2}\ \frac{d\tilde \varphi_1\wedge d\tilde \varphi_2}{\zeta ^2+r^4 \sin ^2(2 \xi )}\,,
\end{aligned}
\ee
where $ds_{\alg{a}}^2 $ is the metric of AdS$_5$.
After computing the matrix $M_{ab}$ and the local Lorentz transformation\footnote{For $32\times 32$ Gamma matrices we find convenient the basis used in~\cite{Arutyunov:2015qva}.} we get that only $F^{(3)}$ and $F^{(5)}$ are non-vanishing
\be
\begin{aligned}
&F^{(3)}=4 r^3  \sin (2 \xi ) d\varphi \wedge d\xi \wedge dr,\\
&F^{(5)}=-2 \zeta(1+*)\left(\frac{  r^3 \sin (2 \xi )\, d\tilde \varphi_1\wedge d\tilde \varphi_2\wedge d\varphi   \wedge d\xi \wedge dr}{\zeta ^2+r^4 \sin ^2(2 \xi )}\right).
\end{aligned}
\ee
Since $\omega$ is non-degenerate on $\tilde{\alg{g}}$ we can relate the above background to a YB deformation of AdS$_5\times$S$^5$, see also section~\ref{sec:YB}. In this particularly simple example the $R$-matrix of the YB model is abelian, and therefore it corresponds  just to a TsT transformation on the sphere, see also~\cite{Osten:2016dvf}.
In fact, consider  the following TsT transformation on AdS$_5\times$S$^5$
\be
\varphi_1\to T(\varphi_1),\qquad
\varphi_2\to \varphi_2-2\eta T(\varphi_1),\qquad
T(\varphi_1)\to \varphi_1,
\ee
which produces the following background\footnote{As a starting point we take the undeformed AdS$_5\times$S$^5$ background as written in~\cite{Arutyunov:2015qva}.}
\be
\begin{aligned}
&ds^2=ds_{\alg{a}}^2+
\frac{r^2 }{1+\eta^2r^4 \sin ^2(2 \xi )}(\cos ^2\xi\, d\varphi_2^2+\sin ^2\xi\, d\varphi_1^2)
+(1-r^2)d\varphi^2+r^2d\xi^2+\frac{dr^2}{1-r^2}\,,\\
&e^\phi=(1+\eta ^2r^4 \sin ^2(2 \xi ))^{-\frac{1}{2}}\,,\qquad
B= -\frac{\eta r^4 \sin ^2(2 \xi ) d \varphi_1\wedge d \varphi_2}{1+\eta^2r^4 \sin ^2(2 \xi )}\,,
\end{aligned}
\ee
for the NSNS sector and
\be
\begin{aligned}
&F^{(3)}=4\eta r^3  \sin (2 \xi ) d\varphi \wedge d\xi \wedge dr,\\
&F^{(5)}=-2 (1+*)\left(\frac{  r^3 \sin (2 \xi )\, d \varphi_1\wedge d \varphi_2\wedge d\varphi   \wedge d\xi \wedge dr}{1+\eta^2r^4 \sin ^2(2 \xi )}\right),
\end{aligned}
\ee
for the RR sector.
To match with the above TsT background we need to implement the field redefinition~\eqref{eq:field-red-DTD-YB} at the level of the DTD background, which in this case just  reduces to $\tilde \varphi_1=\eta^{-1}\varphi_2, \tilde \varphi_2=-\eta^{-1}\varphi_1$ since $\tilde{\alg{g}}$ is abelian.
We find agreement only if we also use the gauge freedom for $B$ to subtract the exact term $\frac{1}{2\eta}d \varphi_1\wedge d \varphi_2$; moreover  we also need to redefine the constant part of the dilaton to reabsorb a factor of $\eta$, which then appears in front of the RR fields.

\subsection{A new example}\label{sec:ex-not-YB}
{Let us now consider the example in (\ref{eq:ex1})}
\begin{equation}
\tilde{\mathfrak g}=\mathrm{span}\{p_1,\,p_2,\,p_3,\,J_{12}\}\,,
\qquad
\tilde{\mathfrak g}^*=\mathrm{span}\{-\tfrac{1}{2}k_1,\,-\tfrac{1}{2}k_2,\,-\tfrac{1}{2}k_3,\,-J_{12}\}
\qquad\omega=k_3\wedge J_{12}\,.
\end{equation}
In this case we have just one isometry of type 1 corresponding to $p_0$, and the isometries of type 2 are $k_3$ and $J_{12}$. Inspired by the parametrisation used in (6.19) of~\cite{Borsato:2016ose} we parametrise\footnote{Even if present, one could remove $k_2$ in $\nu$ by means of a gauge transformation.}
\be
\nu=\tilde \xi\ J_{12}+\tilde r\ k_1+ \tilde x^3\ k_3\,,
\qquad
f=\text{exp}(x^0p_0)\, \text{exp}(\log z D)\,.
\ee
The above is a good parametrisation because it is not possible to remove degrees of freedom by applying gauge transformations. This will be confirmed e.g. by the fact that we get a non-degenerate  metric in target space. We find that the (matrix corresponding to the) operator $\op$ is
\be
\tilde O_{ij}=\left(
\begin{array}{cccc}
 \frac{2}{z^2} & 0 & 0 & 0 \\
 0 & \frac{2}{z^2} & 0 & 2 \tilde{r} \\
 0 & 0 & \frac{2}{z^2} & 2 \zeta  \\
 0 & -2 \tilde{r} & -2 \zeta  & 0 \\
\end{array}
\right)\,,
\ee
which is clearly invertible. We  find the following NSNS sector fields
\be
\begin{aligned}
&ds^2=\frac{-(dx^0)^2+dz^2}{z^2}+d\tilde r^2z^2
+\frac{d\tilde \xi^2}{4 z^2 \left(\zeta
   ^2+\tilde{r}^2\right)}+\frac{\tilde{r}^2 z^2(d\tilde x^3)^2}{\zeta
   ^2+\tilde{r}^2}+ds_{\alg{s}}^2
\,,\\
&e^\phi=\left(\frac{16 \left(\zeta ^2+\tilde{r}^2\right)}{z^4}\right)^{-\frac12}\,,\qquad
B= -\frac{\zeta  d\tilde{\xi}\wedge d\tilde x^3 }{2 \left(\zeta
   ^2+\tilde{r}^2\right)}\,,
\end{aligned}
\ee
where $ds_{\alg{s}}^2$ is the metric on S$^5$.
In the RR sector we have only three-form flux
\be
F^{(3)}=-\frac{8  (dx^0\wedge d\tilde{\xi}\wedge dz)}{z^5}\,.
\ee
According to the discussion in section~\ref{sec:YB} the above background is not related to a YB model by NATD.


\section{Conclusions}
We have argued that DTD models based on supercosets represent a large class of integrable string models which is closed under NATD as well as (certain) deformations. Besides being a useful tool to generate new integrable supergravity backgrounds it would be very interesting if these deformations could be understood on the dual field theory side. In the case when the 2-cocycle is invertible these models are equivalent to YB sigma models, which have been argued to correspond to non-commutative deformations, e.g. \cite{Hashimoto:1999ut,Maldacena:1999mh}, of the field theory {\cite{Matsumoto:2014gwa,vanTongeren:2015uha,vanTongeren:2016eeb}} (see also \cite{Araujo:2017jkb}). This interpretation is consistent with the fact that TsT transformations are special cases of these models \cite{Matsumoto:2014nra,Osten:2016dvf} and this includes the so-called $\beta$ and $\gamma$-deformations which have a known interpretation in $\mathcal N=4$ super Yang-Mills \cite{Leigh:1995ep,Lunin:2005jy,Frolov:2005ty,Frolov:2005iq}. Recently a certain limit of the $\gamma$-deformation has been used to construct a simplified integrable scalar field theory \cite{Gurdogan:2015csr,Gromov:2017cja} and it would be very interesting to explore similar limits of the more general class of deformations considered here to see whether one can learn more about the AdS/CFT duality for those cases.

Another important question is how the DTD model relates to the other known deformations of the $AdS_5\times S^5$ string, i.e. the $\eta$-model with R-matrix solving the modified CYBE \cite{Delduc:2013fga} and the $\lambda$-model \cite{Hollowood:2014qma}. These two deformations are related by Poisson-Lie T-duality and the fact that the latter is Weyl-invariant \cite{Borsato:2016ose} while the former is not \cite{Arutyunov:2015qva,Arutyunov:2015mqj} is explained by the fact that the obstruction to the duality at the quantum level again involves the trace of the structure constants \cite{Tyurin:1995bu}.\footnote{We thank A. Tseytlin for this comment.} The fact that NATD is used also in the construction of the $\lambda$-model suggests that there might be a bigger picture relating it to the DTD construction considered here. In fact this seems to be part of an even bigger picture of general integrable deformations of sigma models where T-duality and its generalizations play a central role, see for example the recent paper \cite{Klimcik:2017ken}.

\section*{Acknowledgements}
{We thank  B. Hoare, S. van Tongeren and A. Torrielli for interesting and useful discussions, and A. Tseytlin for illuminating discussions and comments on the manuscript.}
The work of R.B. was supported by the ERC advanced grant No 341222.
We also thank the Galileo Galilei Institute for Theoretical Physics (GGI) for the hospitality and INFN for partial support during the completion of this work at the program {\it New Developments in $AdS_3/CFT_2$ Holography}.

\vspace{2cm}

\appendix


\section{Useful identities}
A useful identity is
\begin{equation}
\tilde P[\tilde P^Tx,(1-\tilde P)y]=0\,,\qquad\forall x,y\in\mathfrak g
\label{eq:PPT-rel}
\end{equation}
which is easily proven by taking the supertrace with an element of $\mathfrak g$.
We will also need some relations related to the well-known formula for the derivative of the exponential map
\begin{equation}
de^x=e^x\frac{1-e^{-\ad_x}}{\ad_x}dx\,.
\end{equation}
Let $x\in\tilde{\mathfrak g}$ and define a similar looking object $\mu=\tilde P^Te^{-x}\delta e^x$, where $\delta$ is the derivation acting as $\delta(x)=\omega(x)$ on $x\in\tilde{\mathfrak g}$. Note that this derivation is compatible with the Lie bracket due to the 2-cocycle condition (\ref{eq:cocycle2}), and following the same computations needed to prove the identity above, one may show that
\begin{equation}
\mu=\tilde P^T\frac{1-e^{-\ad_x}}{\ad_x}\omega x.
\end{equation}
Taking $y\in\tilde{\mathfrak g}$,  from the definition of $\mu$ we find $\tilde P^T\ad_\mu y=\tilde P^T\Ad_{e^x}^{-1}\delta(\Ad_{e^x}y)-\delta y$ which implies
\begin{equation}
\tilde P^T\ad_\mu\tilde P=\tilde P^T e^{-\ad_x}\omega e^{\ad_x}\tilde P-\omega.
\label{eq:admu}
\end{equation}
Another useful identity valid for the derivative of $\mu$ is
\begin{equation}
d\mu=
\mu e^{-x}de^x
+\delta(e^{-x}de^x)
+\tilde P^Tde^{-x}e^x\mu
=
\tilde P^T(\ad_\mu+\omega)(e^{-x}de^x)\,.
\label{eq:dmu}
\end{equation}
Now, the identity~\eqref{eq:PPT-rel} implies that 
\begin{equation}
\tilde P^T\ad_{\tilde P^T\Ad_{\tilde h}\nu}\tilde P=\tilde P^T\Ad_{\tilde h}\ad_\nu\Ad_{\tilde h}^{-1}\tilde P
=
\tilde P^T\Ad_{\tilde h}\tilde P^T\ad_\nu\tilde P\Ad_{\tilde h}^{-1}\tilde P
\end{equation}
and together with~\eqref{eq:admu} it implies that if we redefine $\nu\to\tilde P^T\left(\Ad_{\tilde h}\nu+\zeta\mu\right)$ as in~\eqref{eq:LocalGtilde} then the operator in~\eqref{eq:def-op} transforms as
\be\label{eq:transfo-op}
\op\rightarrow\tilde P^T\Ad_{\tilde h}\op\Ad_{\tilde h}^{-1}\tilde P.
\ee
Moreover, using~\eqref{eq:dmu} we also find
\begin{equation}\label{eq:transfo-dnu}
d\nu\rightarrow\tilde P^T\Ad_{\tilde h}(d\nu-(\ad_\nu+\zeta\omega)(\tilde h^{-1}d\tilde h)).
\end{equation}

\section{Derivation of the action}\label{app:action}
To derive the action of DTD models we start from the action of a supercoset {sigma model}, see e.g.~\cite{Arutyunov:2009ga}, and we rewrite the group element as $g=\tilde gf$, where $\tilde g\in \tilde G\subset G$. We then gauge the $\tilde G$ symmetry and introduce the gauge fields $\tilde A_i$. If we fix the gauge $\tilde g=1$ we essentially achieve $\tilde g^{-1}d\tilde g\to \tilde A$ when comparing to the initial supercoset action. At this point we add a Lagrange multiplier to impose the flatness of $\tilde A_i$, plus a $\omega$-dependent term which deforms the model
\be\label{eq:action-JAnu}
S=-\frac{T}{2}\int d^2\sigma\left[ \frac{\gamma^{ i j}-\epsilon^{ i j}}{2}\STr\left((\tilde{A}_ i+J_ i)\hat d_f(\tilde{A}_ j+J_ j)\right) 
-\epsilon^{ i j}\Str\left(\nu (\partial_ i\tilde A_ j+\tilde A_ i\tilde A_ j)-\frac{\zeta}{2}\tilde A_ i\omega\tilde A_ j\right) \right].
\ee
Instead of integrating out $\nu$ we integrate out $\tilde A$, so that we obtain the equations of motion
\be\label{eq:eom-Atilde}
P^{ i j}_-\left(\op \tilde A_ j+\partial_ j\nu+\hat d_fJ_ j\right)
+P^{ i j}_+\left(\op^T \tilde A_ j-\partial_ j\nu+\hat d_f^TJ_ j\right)=0,
\ee
where
\be\label{eq:ws-proj}
P^{ i j}_\pm=\frac{\gamma^{ i j}\pm\epsilon^{ i j}}{2},
\ee
are projectors
\be
P^{ i j}_++P^{ i j}_-=\gamma^{ i j},
\qquad
P^{ i l}_\pm P^{\ \ \  j}_{\pm l}=P^{ i j}_\pm,
\qquad
P^{ i l}_\pm P^{\ \ \  j}_{\mp l}=0.
\ee
Here we used also $\gamma^{ i j}=\epsilon^{ ik}\gamma_{kl}\epsilon^{l j}$.
We also define $V_{\pm}^ i\equiv P_\pm^{ i j}V_ j$, and it is useful to remember $P^{ i j}_\pm A_ i B_ j=A^ i_{\mp}\gamma_{ i j}B^ j_{\pm}$. We then solve for $\tilde A_\pm$
\be\label{eq:solApm}
\tilde A_-^{ i}=\op^{-1}\left( -\partial_-^ i\nu-\hat d_fJ_-^ i \right)\,,
\qquad
\tilde A_+^{ i}=\op^{-T}\left( +\partial_+^ i\nu-\hat d_f^TJ_+^ i \right)\,.
\ee
The action on the solutions to the equations of motion is
\be\label{eq:action-Jnu}
\begin{aligned}
S&=-\frac{T}{2}\int d^2\sigma\STr\left[ J_{+ i}\hat d_f J_-^{\  i} +(\partial_{+ i}\nu - \hat d_f^T J_{+ i})\op^{-1}(\partial_-^{ i}\nu+\hat d_f J_-^ i) \right]\\
&=-\frac{T}{2}\int d^2\sigma \frac{\gamma^{ i j}-\epsilon^{ i j}}{2}\STr\left[ J_{ i}\hat d_f J_{ j} +(\partial_{ i}\nu - \hat d_f^T J_{ i})\op^{-1}(\partial_{ j}\nu+\hat d_f J_ j) \right]\,.
\end{aligned}
\ee

\section{Classical integrability}\label{sec:integr}
Here we wish to be more explicit and show that the on-shell equations of DTD models can be recast into the flatness condition for a Lax connection. 
The argument follows the one presented in~\cite{Borsato:2016pas} in the case of DTD of Principal Chiral Models.
First we compute the equations of motion for $f$ and $\nu$, which are obtained by the straightforward variations $\delta_f S$ and $\delta_\nu S$ of the action
\be\label{eq:delta-fnu-S}
\begin{aligned}
\delta_f S
&=+\tfrac{T}{2}\int d^2\sigma \STr\left(f^{-1}\delta f\ 
\mathcal C
\right),
\\
\delta_\nu S&=-\tfrac{T}{2}\int d^2\sigma \STr\left(\delta \nu\ \mathcal F^{\tilde A}\right)
=-\tfrac{T}{2}\int d^2\sigma \STr\left((\Ad_f^{-1}\delta \nu)\mathcal F^A\right),
\end{aligned}
\ee
where we defined
\be\label{eq-on-shell-eqs}
\begin{aligned}
\mathcal C&\equiv\partial_{+i}(\hat d A_-^i)+\partial_{-i}(\hat d^T A_+^i)+[A_{+i},\hat dA_-^i]+[A_{-i},\hat d^TA_{+}^i],\\
\mathcal F^A&\equiv\partial_{+i}A_-^i-\partial_{-i}A_+^i+[ A_{+i}, A_-^i]
=-\epsilon^{ij}(\partial_iA_j+A_iA_j),
\end{aligned}
\ee
and similarly for $\mathcal F^{\tilde A}$. Notice that $P^{(0)}\mathcal C=0$. For convenience we also introduced the (projections of the) field $A_\pm^i\equiv \Ad_f^{-1}(\tilde A_\pm^i+J_\pm^i)$, where $\tilde A^i_\pm$ is given in~\eqref{eq:solApm}. 
On the one hand, imposing the equations of motion $\delta_\nu S=0$ is enough to get $\mathcal F^{A}=0$. Notice that this equation is equivalent to imposing separately $\mathcal F^{\tilde A}=0$ and $\mathcal F^{J}\equiv\partial_{+i}J_-^i-\partial_{-i}J_+^i-[ J_{+i},J_-^i]=0$.
On the other hand, the equations of motion $\delta_f S=0$ imply that $\mathcal C$ vanishes only on a certain subspace of the superalgebra $\alg g$. In fact, in the special case when the whole superalgebra is dualised $\tilde{\alg{g}}=\alg g$, there is no $f$ for which we can compute the variation of the action, and we should find an independent argument to claim that the equation $\mathcal C=0$ holds. We will now show that an appropriate (rotated) projection of $\mathcal C$ by $\tilde P^T$ indeed vanishes without appealing to the equations of motion for $f$.
Consider the equations of motion for $\tilde A_\pm^i$ in~\eqref{eq:eom-Atilde} and let us rewrite them as $\mathcal E_\pm^i-M_\pm^{i\perp}=0$ where
\be
\begin{aligned}
\mathcal E_+^i&\equiv+(\partial_+^i+\ad_{\tilde A_+^i})\nu - \hat d^T_f (J_+^i+\tilde A_+^i)-\zeta\omega \tilde A_+^i,\\
\mathcal E_-^i&\equiv-(\partial_-^i+\ad_{\tilde A_-^i})\nu - \hat d_f (J_-^i+\tilde A_-^i)+\zeta\omega \tilde A_-^i.
\end{aligned}
\ee
Since we choose $M_\pm^{i\perp}$ to take values only in the complement of $\tilde{\alg{g}}^*$, taking $\tilde P^T\mathcal E_\pm^i=0$ gives indeed~\eqref{eq:eom-Atilde}. Clearly $(\partial_{+i}+\ad_{\tilde A_{+i}})(\mathcal E_-^i-M_-^{i\perp})+(\partial_{-i}+\ad_{\tilde A_{-i}})(\mathcal E_+^i-M_+^{i\perp})=0$ is identically true since it just follows from the above equations, and working out all the terms we find
\be
\Ad_f\, \mathcal C=[\nu,\mathcal F^{\tilde A}]+\zeta \omega \mathcal F^{\tilde A}
-(\partial_{-i}+\ad_{\tilde A_{-i}})M_+^{i\perp}-(\partial_{+i}+\ad_{\tilde A_{+i}})M_-^{i\perp}.
\ee
After projecting with $\tilde P^T$ all terms with $M_\pm^{i\perp}$ disappear. The remaining terms on the right-hand-side of the above equation vanish thanks to the flatness of $\tilde A$ $(\mathcal F^{\tilde A}=0)$ implied by the equations of motion for $\nu$. To conclude, we obtain $\tilde P^T(\Ad_f\mathcal C)=0$ as wanted, which together with the equations of motion for $f$ is enough to claim $\mathcal C=0$ on the whole superalgebra.

The on-shell equations $\mathcal F^{A}=0$ and $\mathcal C=0$ formally take the same form as  those for a supercoset, where in that case $A$ is the Maurer-Cartan form, see also~\cite{Bena:2003wd, Arutyunov:2009ga}. Therefore one may follow the derivation done in the case of the supercoset, and find that they are encoded in the flatness condition
\be
\epsilon^{ij}(\partial_i\mathcal L_j+\mathcal L_i\mathcal L_j)=0,
\ee 
for the Lax connection
\be
\mathcal L_i=A^{(0)}_i+z A^{(1)}_i+\frac{1}{2}\left(z^2+z^{-2}\right)A^{(2)}_i+\frac{1}{2}\gamma_{ij}\epsilon^{jk}\left(z^{-2}-z^2\right)A^{(2)}_i+z^{-1}A^{(3)}_i,
\ee
where $z$ is the spectral parameter. The existence of a Lax connection implies the presence of a tower of conserved charges, see e.g.~\cite{Torrielli:2016ufi} for a review. However, differently from the case of the supercoset, now fewer of them can be argued to be local. In fact, thanks to the gauge transformation it is always possible to define
\be
\mathcal L'_i= h\mathcal L_ih^{-1}-\partial_ihh^{-1},
\ee
so that $\mathcal L'_i$ is also flat.
In the case of the supercoset, after noticing that $\mathcal L_i(z=1)=A_i=g^{-1}\partial_ig$, one may choose $h=g$ so that the new Lax connection vanishes at $z=1$ $\mathcal L_i'(z=1)=0$. Expanding around that point one finds
\be
\mathcal{L}'_i(z=1+w)=w\ g\left(A^{(1)}_i-2\gamma_{ij}\epsilon^{jk}A^{(2)}_k-A^{(3)}_i\right)g^{-1}+\mathcal{O}(w^2),
\ee
so that the flatness condition for  $\mathcal L'_i$ at order $w$ implies the conservation $\partial_i\mathcal{A}^i=0$ for the current
\be
\mathcal{A}^i=\epsilon^{ij}g\left(A^{(1)}_j-2\gamma_{jk}\epsilon^{kl}A^{(2)}_l-A^{(3)}_j\right)g^{-1}.
\ee
This is how in the supercoset {case} one can argue from the Lax connection that the isometries corresponding to the superalgebra $\alg g$ correspond to \emph{local} charges. In the case of DTD models $A$ is not of the Maurer-Cartan form, and in general it is not possible to find a group element $h$ for which a gauge-equivalent Lax connection vanishes at $z=1$. With the exception of the  isometries discussed in section~\ref{sec:symm}, we therefore expect that the initial symmetries of the undeformed model are traded for non-local charges.


\bibliographystyle{nb}
\bibliography{biblio}{}

\end{document}